\documentclass[twocolumn,prb,floatfix]{revtex4}

\usepackage[dvips]{graphicx}

\newcommand{\PRL}[1]{Phys. Rev. Lett. {\bf #1}}
\newcommand{\PRB}[1]{Phys. Rev. B {\bf #1}}
\newcommand{\PRE}[1]{Phys. Rev. E {\bf #1}}
\newcommand{\eqref}[1]{\mbox{Eq.\ (\ref{#1})}}
\newcommand{\figref}[1]{\mbox{Fig.\ \ref{#1}}}
\newcommand{\secref}[1]{\mbox{Sec.\ \ref{#1}}}
\newcommand{\refref}[1]{\mbox{Ref.\ \onlinecite{#1}}}
\newcommand{\be}{\begin{equation}}
\newcommand{\ee}{\end{equation}}
\newcommand{\ba}{\begin{eqnarray}}
\newcommand{\ea}{\end{eqnarray}}

\begin{document}

\title{	
Scaling, domains, and states in
the four-dimensional random field Ising magnet
}

\date{August 8, 2002}

\author{A. Alan Middleton}
\affiliation{Department of Physics,
Syracuse University, Syracuse, New York 13244}

\begin{abstract}
The four dimensional Gaussian random field Ising magnet is investigated
numerically at zero temperature, using
samples up to size $64^4$, to test scaling theories
and to investigate the nature of domain walls and the thermodynamic limit.
As the magnetization exponent
$\beta$ is more easily distinguishable from
zero in four dimensions than in three dimensions, these results provide
a useful test of conventional scaling theories.
Results are presented for the critical behavior of the heat capacity,
magnetization, and stiffness.
The fractal dimensions of the domain walls at criticality are estimated.
A notable difference from three dimensions is the
structure of the spin domains:
frozen spins of both signs percolate at a disorder
magnitude less than the value at the
ferromagnetic to paramagnetic transition.
Hence, in the vicinity of the transition, there are two percolating
clusters of opposite spins that are fixed under any boundary conditions.
This structure changes the interpretation of the domain walls
for the four dimensional case.
The scaling of the effect of boundary conditions on the interior spin
configuration is found to be consistent with the domain wall dimension.
There is no evidence of a glassy phase: there appears to be a
single transition from two ferromagnetic states to a single
paramagnetic state, as in three dimensions.
The slowing down of the ground state algorithm is also used to
study this model and the links between combinatorial optimization
and critical behavior.
\end{abstract}

\pacs{75.10.Nr, 75.50.Lk, 02.70.Lq, 02.60.Pn}

\maketitle

\section{Introduction}

As the random field Ising magnet (RFIM) is
a relatively well-studied model of a disordered material,
general questions about thermodynamic
phases and transitions have been addressed using it as a model system.
Experimental studies of random
field Ising magnets are also available for comparison with theoretical
predictions.
The statics of the RFIM have been studied in detail
theoretically, both
analytically and numerically.
It has been proven that there are at least two phases in dimensions
greater than two,\cite{proofs}
scaling arguments have been constructed,\cite{NattermannInYoung}
the replica approach has been applied,\cite{MezardYoungMonasson}
and the model has been analyzed
on hierarchical lattices.\cite{CaoMachta}
The RFIM also has a rich numerical history, including extensive Monte Carlo
simulations\cite{RiegerRFIM,RiegerYoung,NewmanBarkemaRFIM,MachtaNewmanChayes}
and zero temperature ground state studies.\cite{dAuriac,Ogielski,Swiftetal,HartmannNowak,dAuriacSourlas,Sourlas,HartmannYoung,MiddletonFisherRFIM,DukovskiMachta}
Some questions about the model remain unsettled, though, and the
physical picture of excitations is somewhat incomplete. Studying
these properties of the model will be useful in building a more
complete picture, especially when addressing questions about dynamics.

There has been an active discussion about the nature of the
phase diagram for the random field Ising model (RFIM).
One controversy has been whether the transition
from the ferromagnetic phase to paramagnetic phase, which occurs
as the disorder strength or temperature is varied, is continuous
in three dimensions.\cite{RiegerRFIM,dAuriacSourlas,Sourlas,NattermannInYoung}
Recent work\cite{MiddletonFisherRFIM,DukovskiMachta}
provides further strong evidence that
the transition is second order in this case and that previously
derived scaling relations apply. However, as the ratio $\beta/\nu$ of
the order parameter exponent $\beta$ to the correlation length
exponent $\nu$ is very small, some scaling predictions are hard to verify.
It is of interest to pursue this investigation in higher
dimension, where $\beta/\nu$ is larger, to verify the general theoretical
picture suggested for the RFIM in finite dimensions.

\section{Summary of results}

Numerical simulations have been carried out for the Gaussian RFIM on
a simple hypercubic lattice in four dimensions.
The Hamiltonian (for a review of the RFIM see \refref{NattermannInYoung})
is defined over spin configurations $\left\{s_i=\pm 1\right\}$,
\begin{equation}
H = - J\sum_{\left<ij\right>} s_i s_j - \sum_{i}h_i s_i,
\end{equation}
with $\left<ij\right>$ indicating nearest neighbor sites $i,j$
and the random fields $h_i$ are
chosen independently from a Gaussian distribution
with mean zero and variance $h^2$. Here the
energy scale is fixed by setting $J=1$ in the computations,
with temperature $T=0$.
Exact ground states for this Hamiltonian are found using a max-flow
algorithm, as in previous
work.\cite{dAuriac,Ogielski,Swiftetal,HartmannNowak,dAuriacSourlas,Sourlas,HartmannYoung,MiddletonFisherRFIM,DukovskiMachta}

The magnetization is more useful in studying the 4D RFIM
than the 3D RFIM, as the magnetization exponent $\beta$ is
more easily distinguished from zero. The Binder parameter is used to locate
the ferromagnetic-to-paramagnetic
transition relatively precisely at $h_c=4.179(2)$.
A finite-size study of the
magnetization allows the ratio $\beta/\nu$ to be
estimated as $\beta/\nu=0.19(3)$. Besides its relevance to the
magnetization, this ratio is important in studying
the nature of the states and comparing domain wall exponents.

The ground state energies and their dependence on boundary
conditions can be used to study the heat capacity and stiffness
exponents of the RFIM. The stiffness (violation of hyperscaling)
exponent is determined to be $\theta=1.82\pm 0.07$, consistent
with conventional exponent bounds.\cite{SofferSchwartz,FisherRFIM}
Unlike the 3D case,
the value for $\theta$ is numerically distinguishable from $d/2$.
The heat capacity exponent inferred from the
ground state energies is estimated as $\alpha = 0.26\pm 0.05$, apparently
distinct from $\alpha = 0$ and again consistent with the conventional
disorder variant of Widom scaling,\cite{FisherRFIM,VillainRFIM}
$(d-\theta)\nu = 2-\alpha$.

The spatial structure of the spins for the 4D RFIM
is found to differ from the structure for the 3D RFIM,
over the length scales studied.
In the case of three dimensions,
the spins appear to form
a nested sets of domain walls at
criticality.\cite{MiddletonFisherRFIM,VillainRFIM} 
In 4D, the frozen spins (those invariant
under all boundary conditions) percolate
in the ferromagnetic phase.
This implies that domain walls cannot be simply
identified as surfaces between connected sets of same-sign spins.
Simulations show that
the frozen spins percolate at a value $h_p^f= 3.680(5)$.
At a slightly higher value of $h$, $h_p^m$,
the minority spins (frozen spins of a sign opposite to the magnetization)
percolate.
Evidence is given in \secref{sec_perc} that this percolation takes place even
when the spins are coarse grained, with the critical value of $h$
dependent on the scale of the coarse graining.
While this percolation does {\em not} affect thermodynamic quantities
such as the bond part of the mean ground state energy,
$\overline{E}_J$, or the magnetization,
the definition of the domain walls and the
description of the spin-spin correlation function
turns out to not be as straightforward as in the
case of $d=3$.

The qualitative nature of the thermodynamic limit in the 4D RFIM
can be addressed by studying the influence of boundary conditions
on the configuration in a fixed window.
In \secref{sec_states}, the effect of
up ($+$) and down ($-$) boundary conditions at the surface of the
sample are compared with periodic boundary conditions ($P$).
The probability of the interior spins in the $P$ configuration
being identical to either the $+$ or
$-$ configuration approaches $1$, as $L\rightarrow\infty$, for all $h$.
Taking the periodic boundary condition as a generic case, then,
in the large volume limit, the
interior of the ground state configuration is found in
one of the two ferromagnetic states for $h<h_c$ or the paramagnetic
state for $h>h_c$. The probability for the P configuration
to be either $+$ or $-$ in the interior scales in a manner consistent
with the 3D results\cite{MiddletonFisherRFIM}
and the general case where there are few states
in the thermodynamic limit.\cite{AAMstates}

\subsection{Algorithm and error bars}

The variant of the push-relabel algorithm used is the same as
described in Ref.\ \onlinecite{MiddletonFisherRFIM}.
Near criticality, ground states for samples of size $64^4$ were found in
about \mbox{3000 s}
using 1 GHz Pentium III processors. Ground states for smaller
samples were found using a faster, but
less memory-efficient, version of the algorithm; using the same
processors, near-critical
samples of volume $32^4$ were solved in approximately \mbox{60 s}.

Error bars for exponent
values throughout this paper
include both estimated systematic errors due to apparent
finite size effects and errors due to statistical uncertainties;
the error bars represent an estimated range of values in which the
value lies, with high confidence. In contrast, error bars in the figures
for raw data reflect $1\sigma$ statistical
uncertainties computed from the standard deviation,
except for the Binder cumulant,
where the error bars were computed by resampling.
Plots of fitted values, such as estimated peak heights, include
both statistical errors and an error bar that reflects fluctuations
in values that result from varying the degree of the polynomial fit
and the chosen range of the fit.

In some of the plots, exponent values that differ from the ``best''
value from other plots are used to scale the data,
to indicate
that there is some flexibility in the exponent values, depending on the
method. All of the values derived for the exponents from various methods
are consistent with each other to within statistical and
estimated systematic errors.
Table\ \ref{numtable} gives a summary of
the numerical values of the best estimates for the exponents.

\begin{table*}
\caption{Table of numerical estimates for the 4D Gaussian RFIM on the simple
hypercubic lattice.}
\label{numtable}
\begin{tabular}{|ccl|}\hline
Symbol & Value & Definition and data used \\
\hline
$h_c$ & $4.179 \pm 0.002$ & Critical value of the random field for coupling $J=1$.\\
& & The critical point is determined primarily from scaling of magnetization\\
& & distribution (e.g., Binder cumulant as shown in \figref{fig_binder});\\
& & this $h_c$ is consistent with extrapolation in $L$ of the location of peaks in the\\
& & specific heat and the number of operations used to find the ground state\\
& & and the value at which the probability of stiffness being zero is independent of $L$.\\\hline
$h_p^f$ & $3.680 \pm 0.005$ & Value of the random field at which the frozen spins percolate. \secref{sec_perc}.\\\hline
$h_p^m$ & $3.875 \pm 0.005$ & Value of the random field at which the minority spins percolate. See \secref{sec_perc} and \figref{fig_percb}.\\\hline
$\beta/\nu$ & $0.19 \pm 0.03$ &
Ratio of magnetization exponent $\beta$ to correlation length exponent $\nu$.\\
& & Determined from the scaling of
$|m|$ vs. $L$ at criticality for $12 \le L \le 64$.\\
& & See \figref{fig_betanu}, \figref{fig_mag2}
and \secref{sec_mag}.\\\hline
$\alpha/\nu$ & $0.31 \pm 0.04$ &Heat capacity
exponent $\alpha$ divided by $\nu$.\\
& & Found from peaks $C_{\rm max}(L)$,
computed from the derivative of fit to $\overline{E}_J(h,L)$.\\
& & See Figs.\ \ref{fig_EJ} and \ref{fig_cmax}.\\\hline
$(\alpha -1)/\nu$ & $-0.94 \pm 0.06$ &Combination of heat capacity
exponent $\alpha$ and $\nu$.\\
& & Found from fit to power law for the discrete
estimate of $d\overline{E}_J/d\ln(L)$ evaluated at peak of $C$.\\\hline
$\nu$ & $0.82 \pm 0.06$ & Correlation length exponent.\\
& & Jointly estimated from magnetization scaling, $\alpha/\nu$ and $(\alpha-1)/\nu$, and the scaling of the stiffness.\\
& & with $L$. Consistent with the scaling of the width of the number of algorithm operations.\\\hline
$\beta$ & $0.16 \pm 0.03$ & Magnetization exponent, found from $\beta/\nu$ and $\nu$.\\\hline
$\alpha$ & $0.26 \pm 0.05$ & Heat capacity exponent,
found from $\alpha/\nu$ and $\nu$. \\\hline
$\theta$ & $1.82 \pm 0.07$ & 
Violation of
hyperscaling or the scaling of the stiffness at $h_c$.
.\\
& & Found from scaling of stiffness with $L$ and $h-h_c$,
see \secref{sec_stiffwalls} and \figref{fig_stiffness}.\\\hline
$d_s$ & $3.94 \pm 0.06$ &
Fractal dimension of connected domain wall at $h=h_c$.\\
& & Note that the result is indistinguishable from $d=4$.\\
& & See \figref{fig_dims} and \secref{sec_stiffwalls}.\\\hline
$d_I$ & $3.20 \pm 0.12$ & Incongruent fractal dimension
of domain wall at criticality.\\
& & Box counting of incongruent volumes (disconnected wall).
See \secref{sec_stiffwalls}.\\
& & Consistent with scaling of state overlap
probabilities shown in \figref{fig_Pscale} and \figref{fig_Pmax}.\\\hline
$d_J$ & $2.94 \pm 0.12$ & Energy ``fractal dimension'' at $h=h_c$.
 Found from the exchange part, $\Sigma_J$, of the stiffness.\\
& & See \figref{fig_dims} and \secref{sec_stiffwalls}.\\\hline
\end{tabular}
\end{table*}

\section{Magnetization}
\label{sec_mag}

\begin{figure}
\centering
\includegraphics[width=8.5cm]
{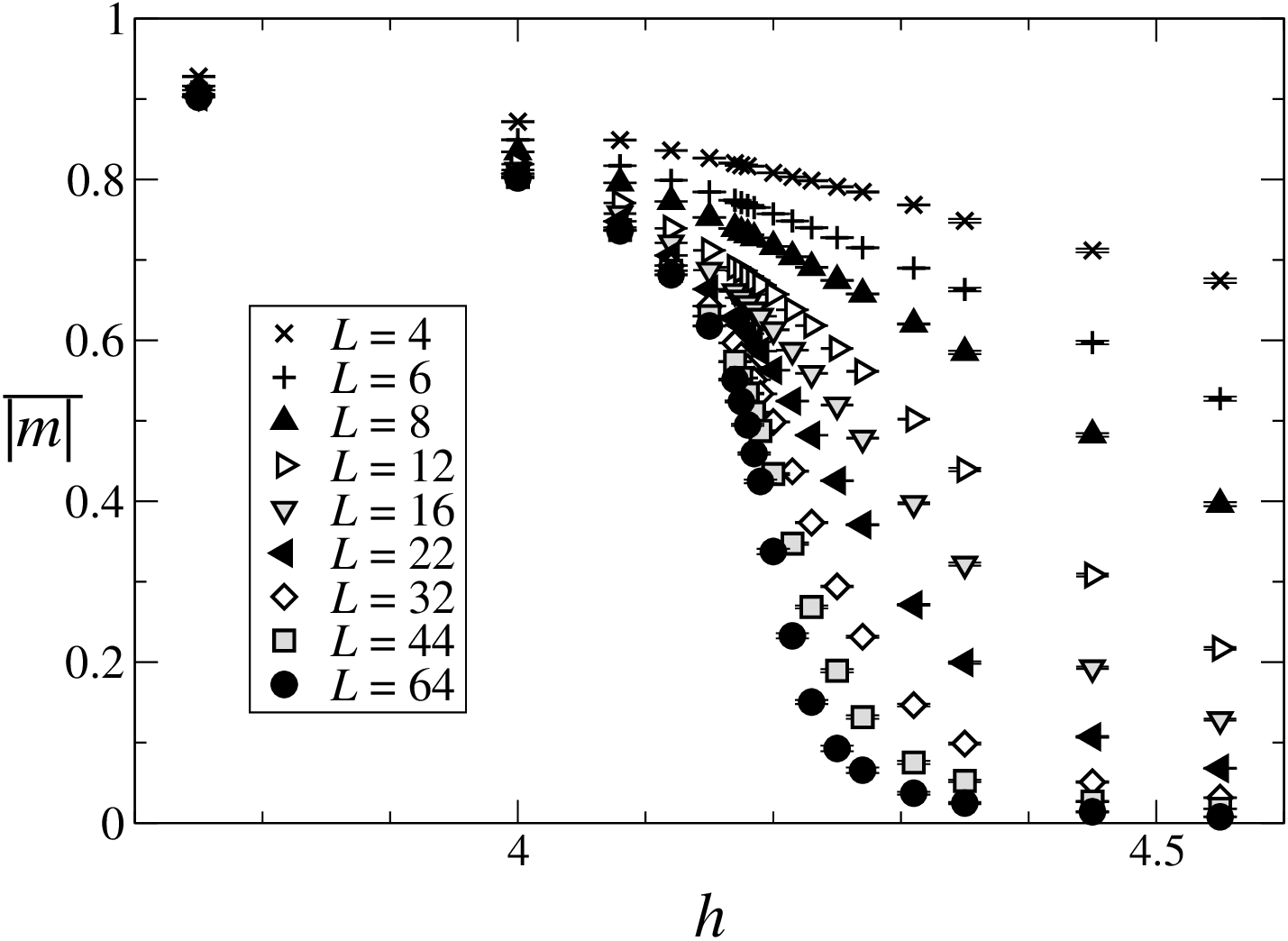}
\caption{
Plot of the sample average of the absolute value of the
magnetization, $\overline{|m|}$, of the 4D RFIM
as a function of disorder $h$ and
system size $L$, for periodic boundary conditions.
}
\label{fig_mag1}
\end{figure}

As the exponent $\beta$ is more readily determined in the
4D RFIM, compared with the 3D case, it is useful in 4D to study the
magnetization as a first guide
to the critical behavior and to locate the transition.
The mean value of the absolute value of the magnetization is defined as
\be
\overline{|m|} = N^{-1}\overline{|\sum_i s_i|},
\ee
where the overline indicates an average over samples of volume
$N=L^4$.
The magnetization is directly computed for each sample from the
ground state with periodic boundary conditions.
The dependence of the sample-averaged
magnetization on disorder $h$ is plotted in
\figref{fig_mag1}.

\subsection{Binder cumulant and $h_c$}

\begin{figure}
\centering
\includegraphics[width=8.5cm]
{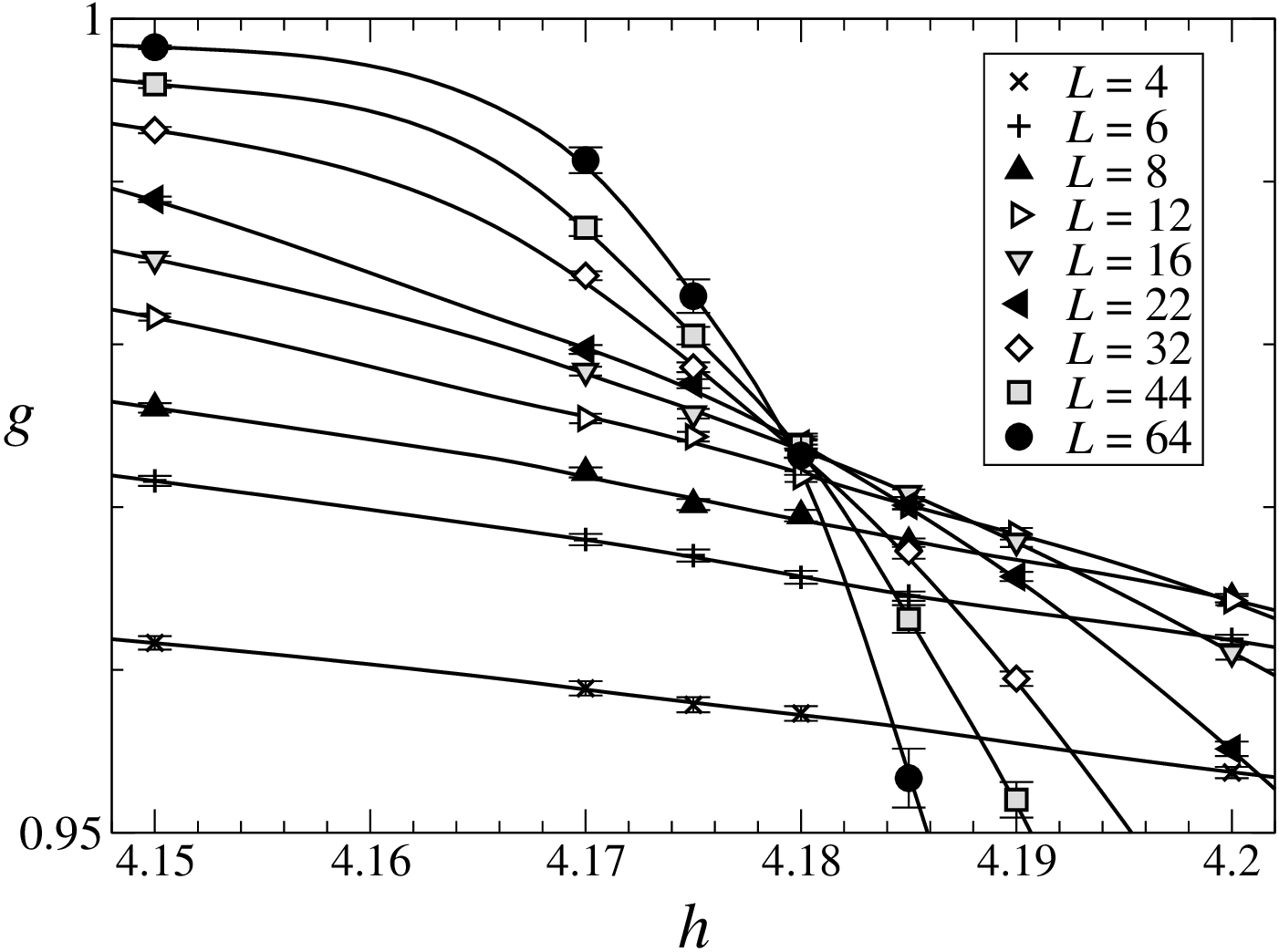}
\caption{
Plot of the Binder cumulant $(3-\overline{m^4}/\overline{m^2})/2$
as a function of $h$ for various $L$.
The curves are smoothed spline fits to indicate the trends.
This plot is used to determine the
location of the transition, $h_c=4.179(2)$.
}
\label{fig_binder}
\end{figure}

One method for determining the value of $h_c$ is to use the
Binder cumulant. The value of the cumulant,
$g=(3-\overline{m^4}/\overline{m^2})/2$, should be $g=1$ in the
ferromagnetic phase and should take on the value $g=0$ in the
paramagnetic phase. The fixed point $h_c$ is found by the intersection
of the $g(h)$ curves for various $L$.
Some caution should be used with this method, as the magnetization
exponent is small, so that the sample distribution of $m$ is bimodal
for even large samples near the transition. The assumptions of Gaussian
behavior about the mean in the paramagnetic phase are difficult to
achieve. Nonetheless, the plots of $g(h)$ show a consistent behavior
that indicates the finite-size trends in the data for the magnetization.
The plot of $g(h)$ for $L=4$ through $L=64$ is shown in \figref{fig_binder}.
For smaller system sizes, about $5\times 10^4$ ground states were found;
for the $L=64$ systems near $h_c$, about $5\times 10^3$ ground states were
computed.
The apparent intersection point is $g_c \approx 0.975$, but
this quantity likely has not converged to its scaling value.
The location of the transition can be assigned with more certainty
to the range $h_c = 4.179\pm0.002$. This is an acceptable value
for the set of lengths used here and likely will continue to hold for
scales that are somewhat larger. It is also quite consistent with the
scaling of the stiffness, location of the specific heat peaks, and the
algorithmic slowing down discussed in the other sections of this paper.

\subsection{Magnetization}

\begin{figure}
\centering
\includegraphics[width=8.5cm]
{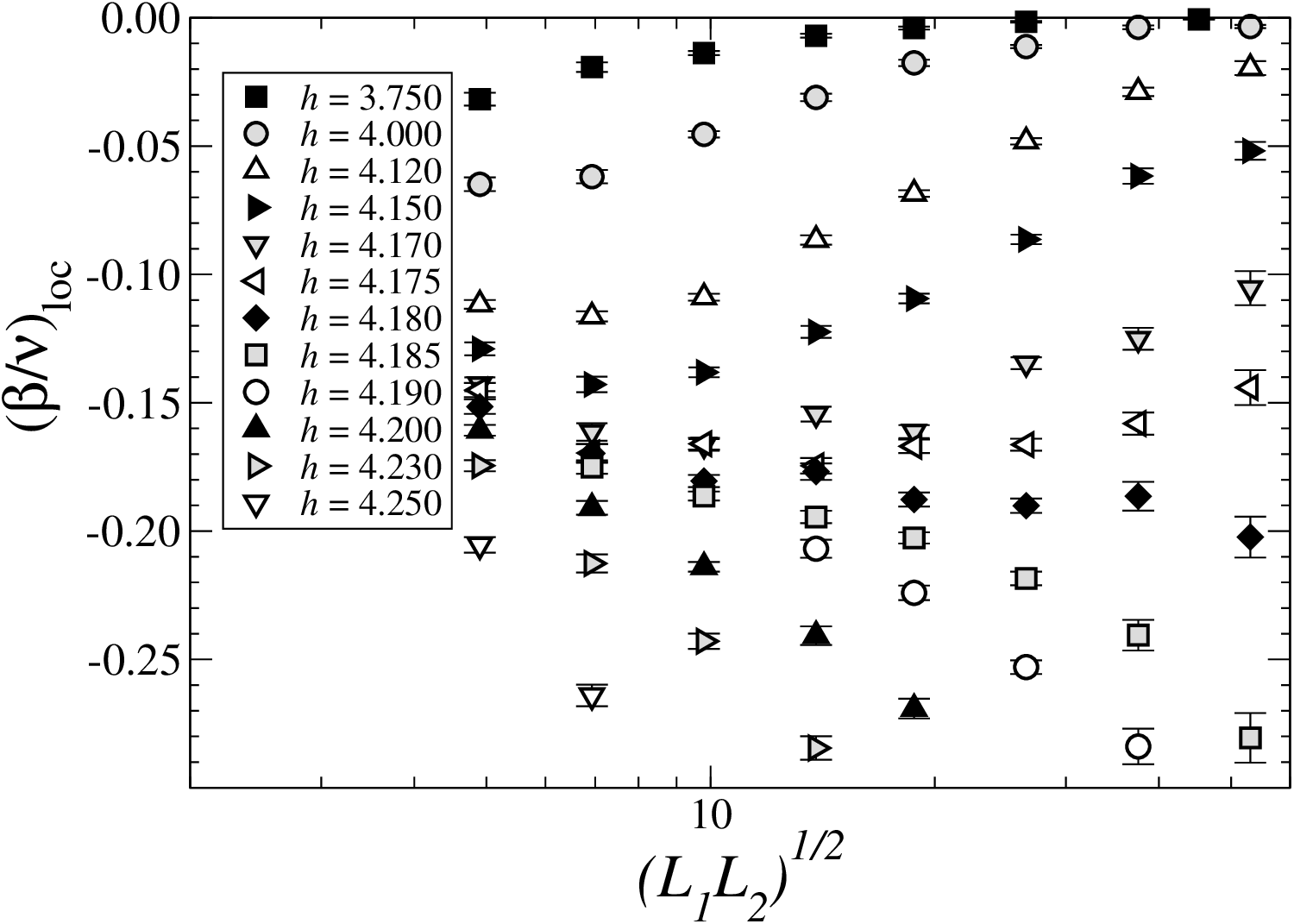}
\caption{
Plot of estimates for $\beta/\nu$ found from the discrete derivative
$(\beta/\nu)_{\rm loc} = \ln(\overline{|m|}(h, L_2)/\overline{|m|}(h, L_1))/
 \ln(L_2/L_1)$. The apparent
convergence to a uniform value for $h \approx 4.18$
implies that the value $\beta/\nu = 0.19(3)$ accurately describes
the effective critical behavior for $L\approx 12 \rightarrow 64$.
}
\label{fig_betanu}
\end{figure}

\begin{figure}
\centering
\includegraphics[width=8.5cm]
{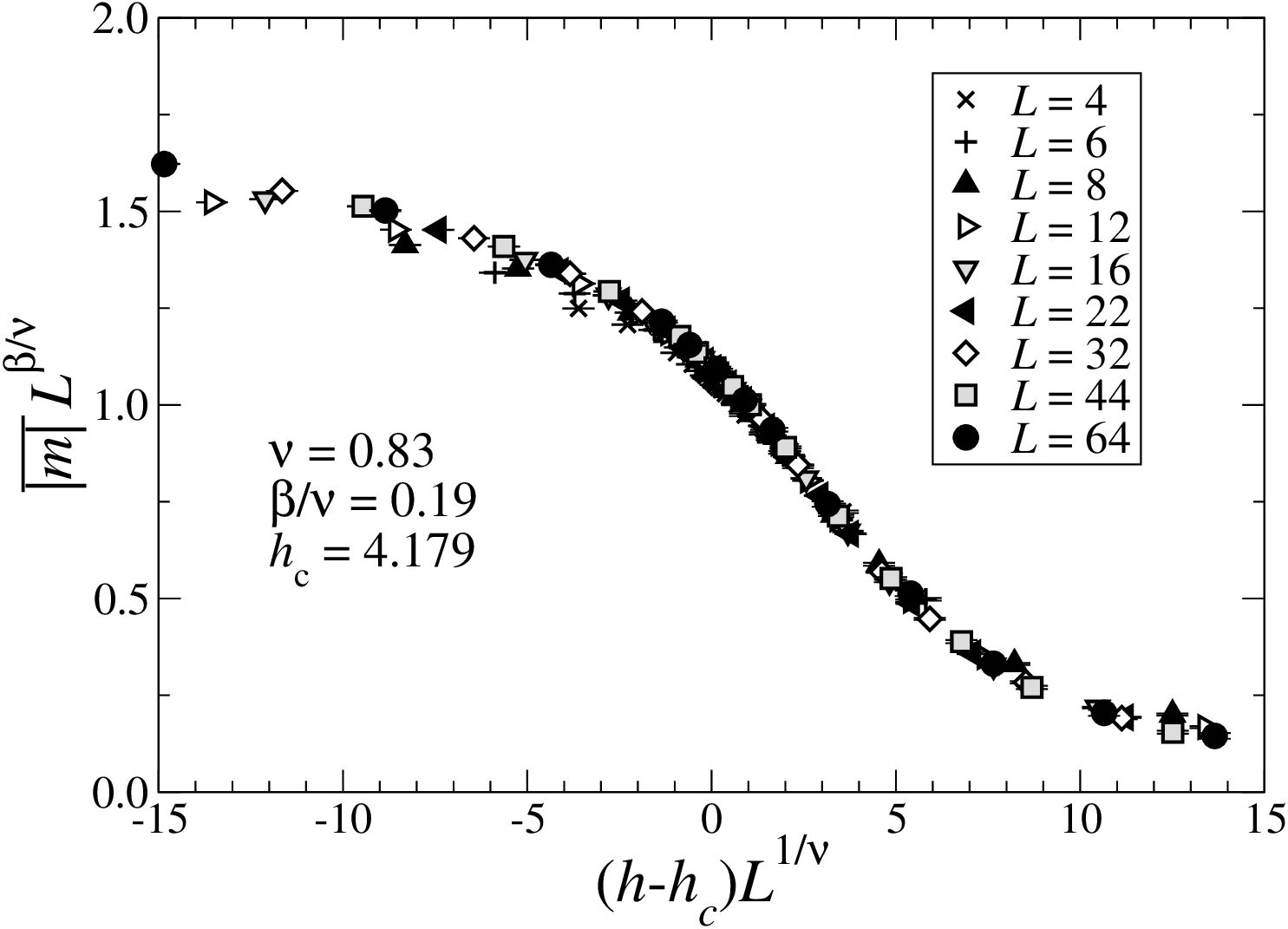}
\caption{
Scaled magnetization
$\overline{|m|}L^{\beta/\nu}$ as a function of scaled disorder
$(h - h_c)L^{1/\nu}$ for $\beta/\nu = 0.19$ and $\nu = 0.83$.
}
\label{fig_mag2}
\end{figure}

The ratio of the magnetization exponent $\beta$ to the correlation
exponent length $\nu$ is computed from the effective finite size
exponent for the magnetization. Given standard finite size scaling,
the magnetization at the transition will scale as
$m(h_c) \sim L^{-\beta/\nu}$. The local exponent $\beta/\nu$ is
found by the discrete derivative of $\ln(\overline{|m|})$ with
respect to $\ln(L)$. The results of this computation are
plotted in \figref{fig_betanu}. This evaluation gives a location
for the transition that is consistent, but slightly less precise,
than the Binder cumulant analysis. The value of the local exponent
that is most consistent with a constant value gives the estimate
\be
\beta/\nu=0.19\pm 0.03.
\ee
In \figref{fig_mag2}, the scaled magnetization (for the same samples
used to compute $g(h)$) is plotted as a function
of scaled distance to the transition, in agreement with the finite-size
scaling form
\be
\overline{|m|}=L^{-\beta/\nu}f_m[(h-h_c)L^{1/\nu}],
\ee
where the value $\nu = 0.83$ gives the best scaling collapse, with
fixed $\beta/\nu = 0.19$ and $h_c=4.179$. The value of $\nu$ in
\mbox{Table\ \ref{numtable}} indicates the range of values found by distinct
estimates; there is no clear best measurement of $\nu$ in the data.

\subsection{Fluctuations in the magnetization}
\begin{figure}
\centering
\includegraphics[width=8.5cm]
{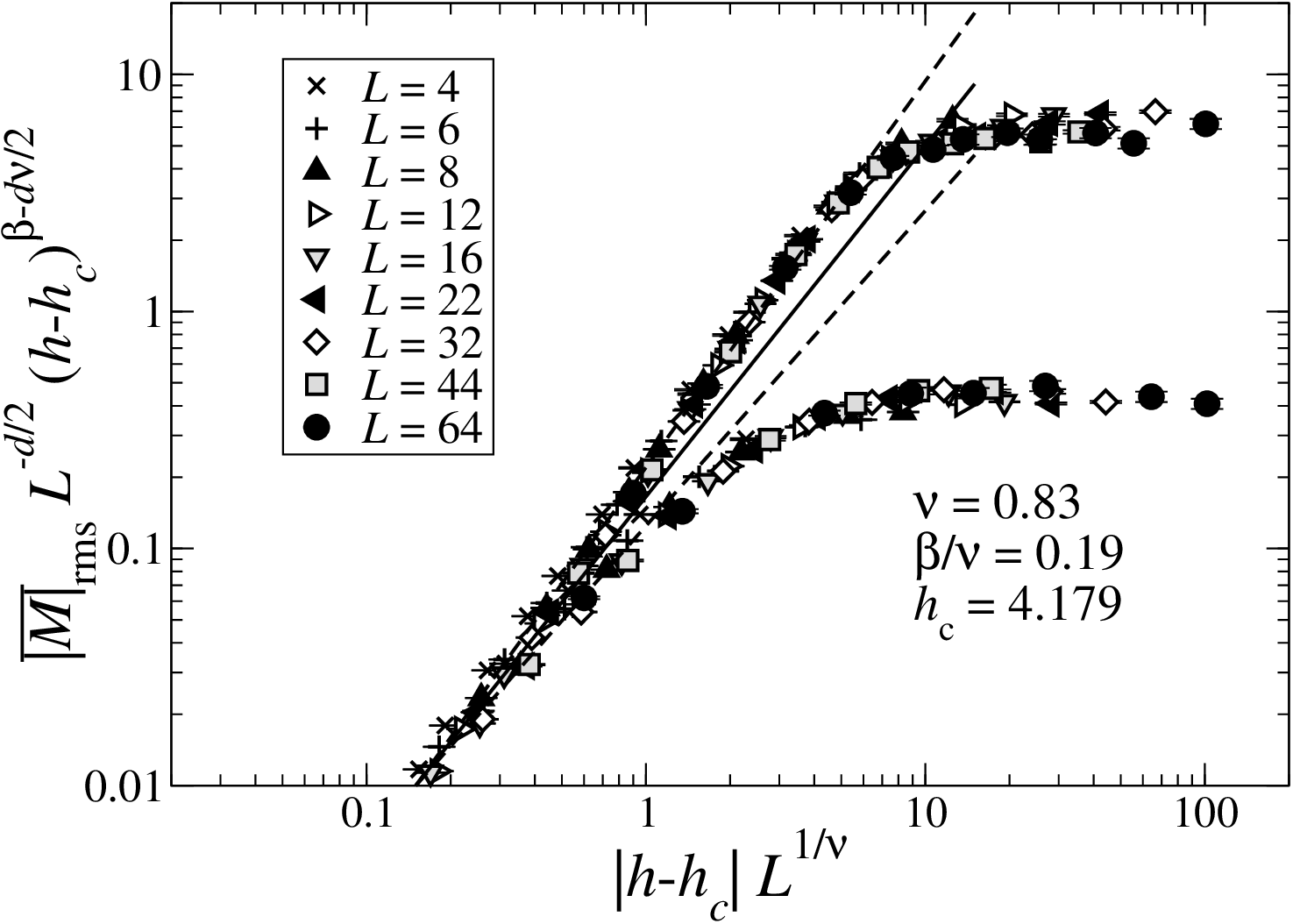}
\caption{Plot of the scaled fluctuations in the magnitude of the magnetization
$|M|$. The scaling variable $|h - h_c|L^{1/\nu}$ on the
horizontal axis is the
scaled distance to the critical point while the vertical axis variable,
$\overline{|M|}_{\rm rms}L^{-d/2}(h - h_c)^{d\nu/2-\beta}$ is
the magnitude of the fluctuation in the magnetization, normalized
by the expected number of correlation volumes and the magnetization of
the correlation volumes of linear size $\xi \sim (h - h_c)^{-\nu}$.
The approach to a constant value at large scaling variable is consistent
with independently oriented regions for $h > h_c$ (upper branch)
and magnetization fluctuations about the ferromagnetic state over regions
of size $\xi$ when $h < h_c$ (lower branch.)
The solid line represents a
power law exponent $-\beta+d\nu/2 \approx 1.48$, while the upper (lower)
dashed line has slope $1.63$ ($1.33$.)
}
\label{fig_fluctmag}
\end{figure}

In addition to the scaling of the mean magnetization, the fluctuations
in the magnetization can be checked for consistency with
finite-size scaling.
The sample-to-sample fluctuations $\Delta_M$ in
the total magnetization $|M| = N {|m|} = \sum_i s_i$ can be
estimated, using the number of independent volumes
and the fluctuations in the magnetization
of such volumes.
Defining the finite size scaling variable $x=(h-h_c)L^{1/\nu}$,
at large $|x|$, the relevant volumes are of size $\xi\sim (h-h_c)^{-\nu}$,
while at small $|x|$, the volume is finite-size limited.
The fluctuations in the magnetization over a volume $\ell^d$ are
$M_\ell \sim \ell^{d-\beta/\nu}$ and the number of
such volumes is $n_\ell \sim (L/\ell)^d$. Taking fluctuations
over each volume to be independent, one can write a version of the
scaling as
\be
\Delta_M \sim M_\ell \sqrt{n_\ell} \sim L^{d/2}(h-h_c)^{d\nu/2-\beta}
f_\Delta(x),
\ee
where in the limit of large $|x|$, $f_\Delta$ approaches a constant whose
value depends on the sign of $|x|$. For small values of $|x|$,
$f_\Delta(x)\sim |x|^{-\beta+d\nu/2}$.
This scaling form is verified by the
data displayed in \figref{fig_fluctmag}.
The data at small $|x|$ is roughly consistent with the range of power laws
$-\beta+d\nu/2 \approx 1.48\pm0.15$ plotted in \figref{fig_fluctmag}
(this comparison is rather sensitive to the location of $h_c$
and the value of $\nu$.) At large $|x|$,
$\Delta_ML^{-d/2}(h-h_c)^{\beta-d\nu/2}$ approaches a constant.

\section{Heat capacity}

The specific heat of the 3D RFIM is a quantity that can be measured
experimentally, directly\cite{Satooka1998,BirgeneauPRL95}
or indirectly.\cite{BirgeneauPRL95,JaccarinoBiref}
The divergence of the specific heat has also been
estimated numerically, though not all estimates agree and the experimental
situation is unclear. Because of these discrepancies, it is useful to
also study this quantity in the case of four dimensions, to check the
validity of the standard scaling picture.

The heat capacity can be estimated using ground state calculations and
applying thermodynamic relations employed by Hartmann and
Young.\cite{HartmannYoung} This approach was also applied in
Refs.\ \onlinecite{MiddletonFisherRFIM}
and \onlinecite{Hartmann4d}. The method relies on studying the
singularities in the bond energy density
\be
E_J=L^{-d}\sum_{\left<ij\right>}s_is_j.
\ee
This bond energy density is the first derivative
$\partial E/\partial J$ of the ground state energy with respect
to $h$ (equivalently, up to constants, with respect to $J$.)
The derivative of the sample
averaged quantity $\overline{E}_J$ with respect to $h$ then gives
the second derivative with respect to $h$ of the total energy and
thus the sample-averaged heat capacity $C$.
The singularities in $C$ can also be studied by computing
the singular part of $\overline{E}_J$, as $\overline{E}_J$
is just the integral of
$C$ with respect to $h$.
The finite-size scaling for the singular
part of the specific heat $C_s$ is
\be
C_s \sim L^{\alpha/\nu}\tilde{C}[(h-h_c)L^{1/\nu}],
\ee
while the scaling for the leading part (through the first
singular term) of the sample averaged
bond energy at $h=h_c$ is
\be
\overline{E}_{J,s}(L,h=h_c) = c_1+c_2L^{(\alpha-1)/\nu},
\ee
with $c_1$ and $c_2$ constants.

\begin{figure}
\centering
\includegraphics[width=8.5cm]
{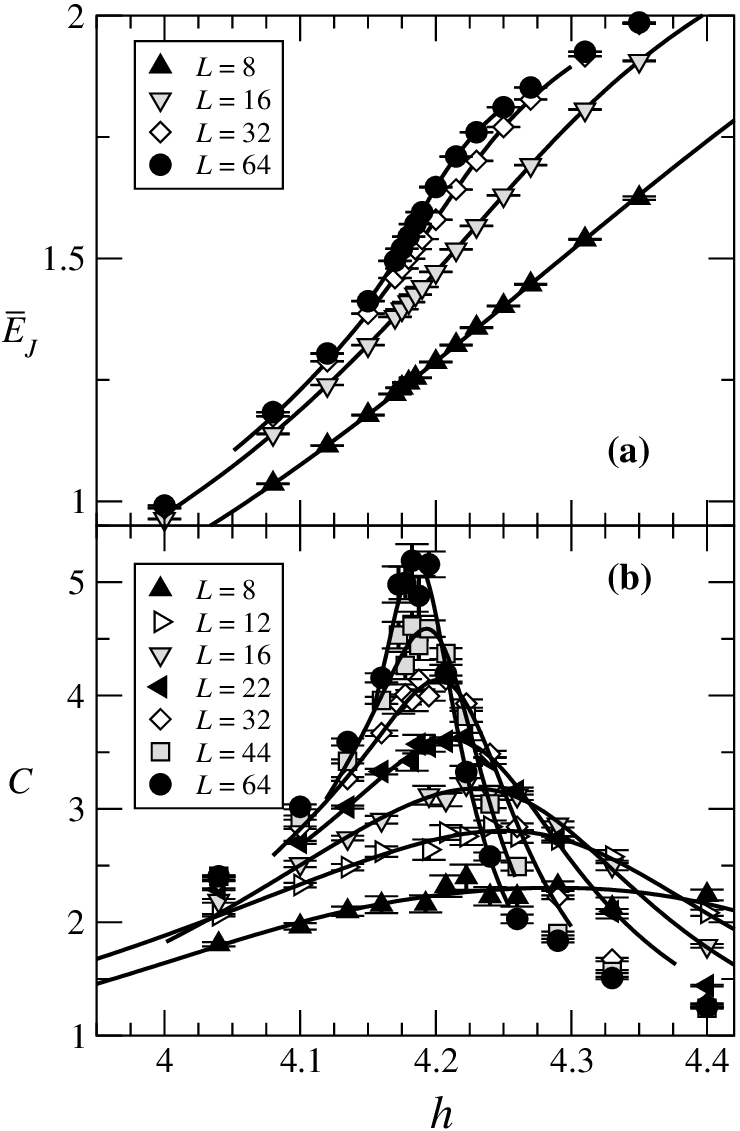}
\caption{(a) The points show the computed dependence of $\overline{E}_J$,
the number density of broken bonds,
on $h$, for $L=4\ldots64$ (not all $L$ values are included, for clarity.) Fits to cubics for the inverse function $h(E_J)$
are shown.
(b) Estimated heat capacity $d\overline{E}_J/dh$, derived from the differences
between the $\overline{E}_J$ values.
Solid lines show the derivatives of the fits
to $\overline{E}_J$ defined in the text.
These derivatives are used to estimate the
heights of the peaks in the specific heat, $C_{\rm max}(L)$.
}
\label{fig_EJ}
\end{figure}

The data analysis is based upon direct fits using $\overline{E}_J$.
This approach avoids complications that arise in computing
the uncertainties when fitting to
finite-differenced estimates for $C$, but is otherwise equivalent to
fitting to such finite differences.
The fit was a least squares fit of a cubic to $h(\overline{E}_J)$ for
fixed values of $L$. This fit to the inverse function was more
stable than fitting to $\overline{E}_J(h)$. The fit function was then
inverted to give the estimate for $\overline{E}_J(h)$.
The maximum slope of this estimated function
is in turn used to estimate the peak in $C(h)$ for each $L$.
The uncertainties at any point, especially when determining
the peak value $C_{\rm max}(L)$, in the analysis
can be estimated using a bootstrap
technique (resampling the data.)
The data for $\overline{E}_J$ are
plotted directly in part (a) of \figref{fig_EJ}.
The samples used were the same as used for the magnetization and Binder
cumulant analysis.
The derivatives of the fit are plotted in part (b) of this
figure and compared with the heat capacity values determined by finite
differencing. Note that the finite differenced values are relatively
noisy due to the differentiation. This apparent noise can be reduced by
less refinement in the values of $h$ sampled, but this
would reduce the resolution in $C(h)$ and the location of the peak in $C$.
By directly fitting to $\overline{E}_J$
rather than the finite differences, this
complication is reduced (but could be managed with appropriate care
in the error analysis.)

\begin{figure}
\centering
\includegraphics[width=8.5cm]
{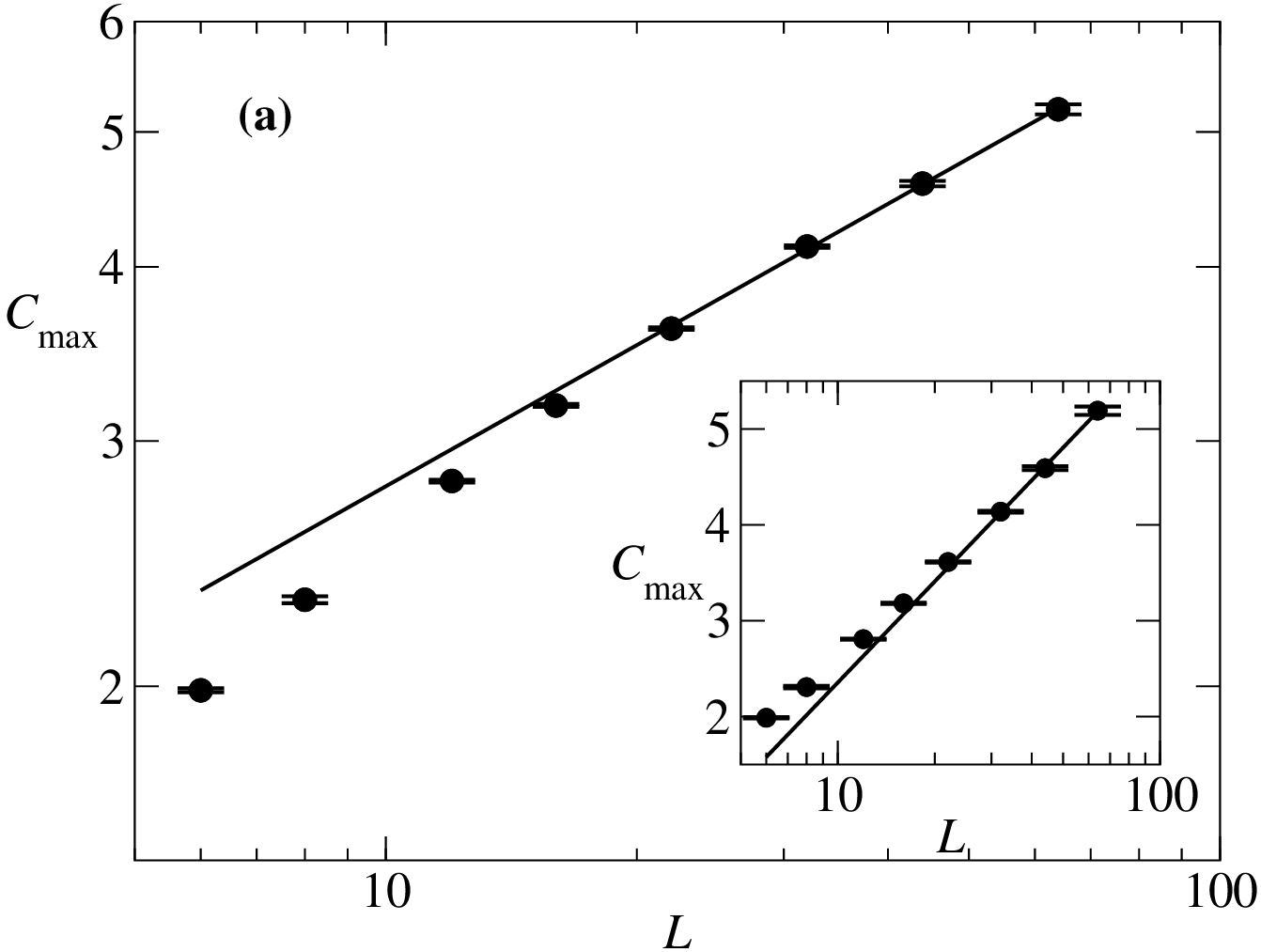}
\centering
\includegraphics[width=8.5cm]
{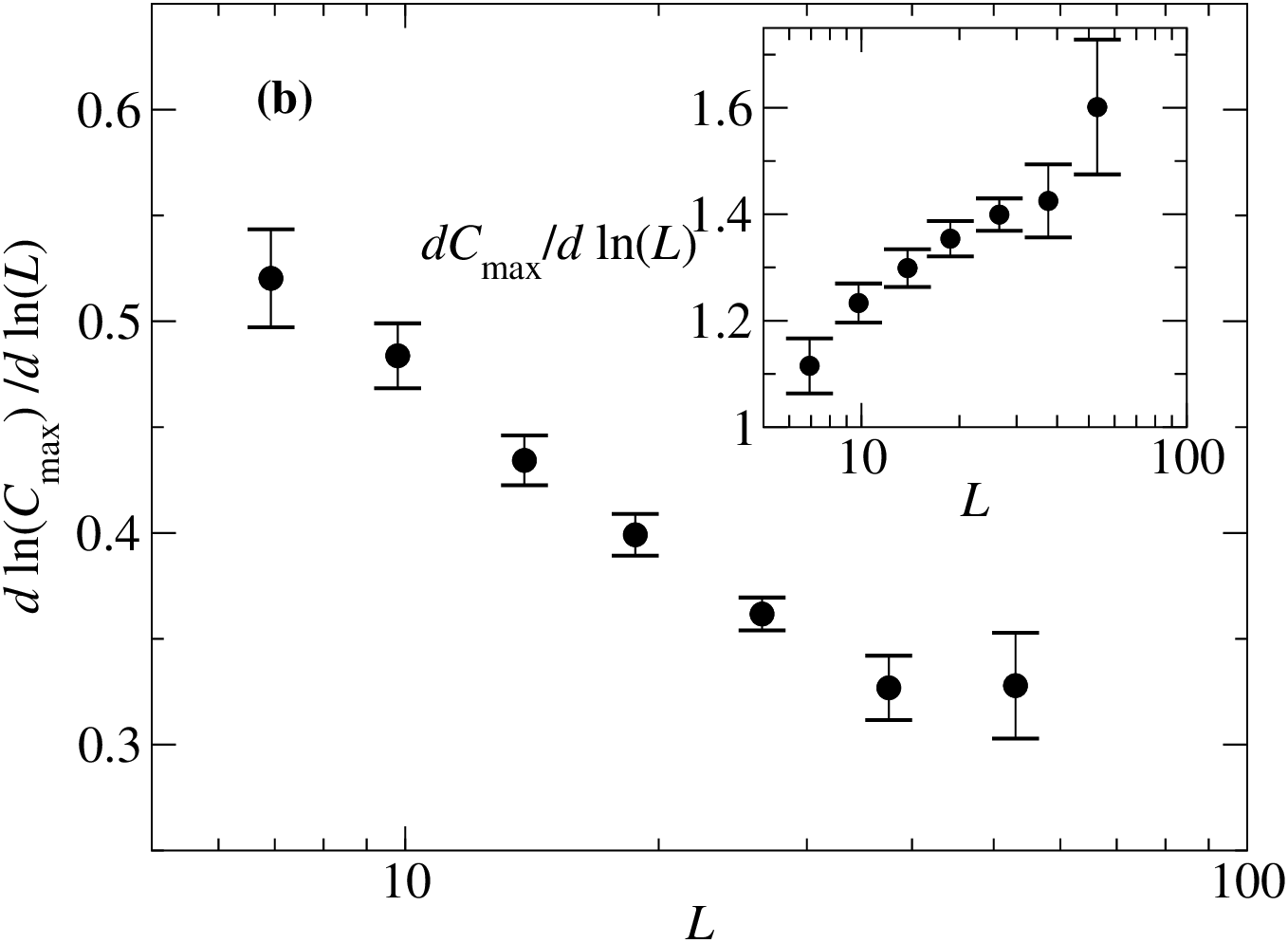}
\caption{
(a) Plot of $C_{\rm max}(L)$ vs. L. The solid line is a fit to the
finite size scaling form $C_{\rm max}(L) \sim L^{-\alpha/\nu}$, with
$\alpha/\nu=0.31\pm 0.04$. The inset shows a semi-log plot
with a least-squares fit
to the last 3 data points ($L\ge 32$.)
(b) Local derivatives of the plots in (a).
}
\label{fig_cmax}
\end{figure}

The estimates for the maximum values of the heat capacity are plotted
as a function of $L$ in \figref{fig_cmax}(a). The relatively precise data
are not consistently fit by a power law until $L>16$. The fit for these
values gives $\alpha/\nu=0.31\pm0.04$, where the errors are purely
statistical. Given the short range of the fit (from $L=22$ to $L=64$),
one must allow for the possibility of corrections to scaling giving a
different value at larger system sizes (possibly slightly lower.)
A fit of these data to $C_{\rm max}\sim\ln(L)$ is less successful, however (see
the inset in \figref{fig_cmax}(a).) As always, it is difficult to distinguish
a logarithmic behavior, suggested for $C_{\rm max}$
in \refref{Hartmann4d}, from
a small-power-law behavior. In \figref{fig_cmax}(b), the local discrete
derivatives are plotted for the cases where the behavior should be
a power law (main part of the figure) and logarithmic (inset.) The
power law does seem to be more consistent with a convergence to a fixed
slope for this range of sizes. Though the fits are not definitive,
the fitted power law behavior
is most consistent with scaling relations and other
data and does seem to explain the computed singularity in the
bond part of the energy.

The specific heat can also be used to infer $\nu$. This can be done
directly through scaling the widths of the peaks in $C$, but a more robust
procedure was to use the indirect procedure
of fitting $\overline{E}_J$, which, being the integral over $h$ of $C$,
incorporates the width of the peak in $C$.
The quantity $(\alpha-1)/\nu$ was found by using the fitted value
of $\overline{E}_J$ at $h_c$.
The derivative of $\overline{E}_J(h_c)$ with respect to $\ln(L)$ gives
the power law $(\alpha-1)/\nu=-0.94\pm 0.06$. With the value for $\alpha/\nu$,
this gives the estimate $\nu=0.80\pm 0.06$.

\section{Stiffness \& domain walls}
\label{sec_stiffwalls}

The nature of responses to external perturbations is used to
characterize distinct phases, in general. One of the more important
responses to study is the response
to changes in the boundary conditions.
For example,
ferromagnetic phases in pure materials can be identified due to the
finite energy density of domain walls induced by twisted boundary conditions.
The application of twisted boundary conditions to stiffness
and domain walls to disordered systems was introduced for spin glasses
by
McMillan\cite{McMillanDWRG} and
Bray and Moore.\cite{BrayMooreLCRG}
Stiffness and domain walls were studied for the 3D RFIM in
\refref{MiddletonFisherRFIM}. The approach taken
quantities studied here for the 4D RFIM
are the same, though the results are somewhat
distinct in flavor from the 3D results.

Measuring the stiffness quantifies the change in energy due to a change
in boundary conditions. The symmetrized stiffness is defined as
\be
\Sigma = (E_{+-}+E_{-+}-E_{++}-E_{--})/2,
\ee
where $E_{ab}$ is the ground state energy for boundary spins fixed to
be $a$ at one end of the sample and $b$ at the other end of the
sample (periodic boundary
conditions are used in the other $d-1$ dimensions).
This definition minimizes the effects of surface terms and has the
value $\Sigma=0$ if the two ends of the sample are ``decoupled'', with
the effect of the boundary conditions penetrating only a finite distance
into the sample. The value $\Sigma$ will be zero
with high probability in the paramagnetic
phase, for large samples, and is expected to scale as $L^{d-1}$ in
the ferromagnetic phase, for fixed $h$.

\subsection{Stiffness at criticality}

The sample averaged stiffness $\overline{\Sigma}$ is a quantity that
is useful for investigating scaling and the order of the transition.
Near a second order transition, the average stiffness scales
with a characteristic scale $L^{\theta}$, where $\theta$ is the ``violation
of hyperscaling'' or stiffness exponent.
The natural scaling assumption is that this stiffness varies
over a scale given by the reduced disorder, giving
\be
\overline{\Sigma} \approx \overline{C} L^{\theta}
{\cal S}[L^{1/\nu}(h-h_c)K],
\label{eq_stiff}
\ee
with $\overline{C}$ and $K$ nonuniversal constants and ${\cal S}$
a function dependent on the shape of the sample.
Another characterization of the distribution of stiffness over samples
is $P_0(h,L)$, which is the probability that the stiffness will be
exactly zero. As the distribution of the stiffness can be scaled at
the critical point, with $\Sigma=0$ invariant under rescaling of $\Sigma$
by $L^\theta$, $P_0(h_c,L)$ approaches a constant as $L\rightarrow\infty$,
with the asymptotic value set by sample shape, disorder distribution, and
lattice type. This convergence to a constant was
used in \refref{MiddletonFisherRFIM} to locate $h_c$ for the 3D RFIM.

\begin{figure}
\centering
\includegraphics[width=8.5cm]
{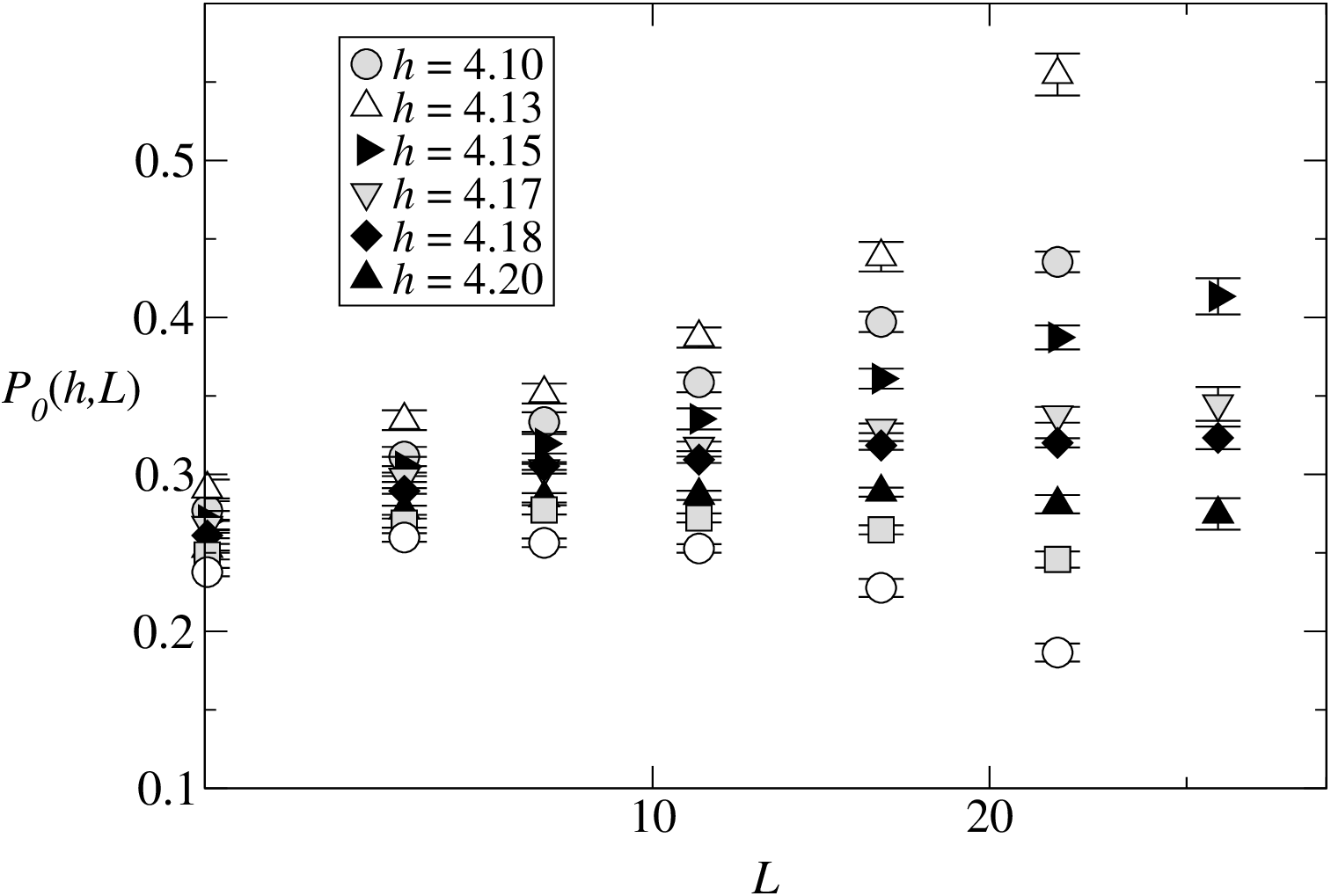}
\caption{
Plot of the probability of zero stiffness $\Sigma$.
The samples have a cross section
volume of $L^3$ with a distance of $3L$ between the controlled
faces. The probability is constant to within numerical errors for
$h \approx 4.18$.
}
\label{fig_P0}
\end{figure}

The probability of zero stiffness $P_0$ is plotted in \figref{fig_P0},
for samples of shape $3L\times L^3$. Less anisotropic samples had a very
small value of $P_0$ and therefore had more statistical error. As the
running time for a given $L$ is larger and the ground states for four
different boundary conditions were computed, fewer samples were studied here
than in the magnetization and energy study.
For $L=32$, up to $5\times 10^3$ realizations were studied, while for
the smallest samples, $E_J$ was calculated for
about $5\times 10^4$ samples.
The estimates plotted are consistent with $P_0$ approximately constant in $L$
for $h_c\approx 4.18$. This is in accord with other estimates of $h_c$,
though the uncertainty in using this plot to determine $h_c$ is somewhat
larger than from other methods.

\begin{figure}
\centering
\includegraphics[width=8.5cm]
{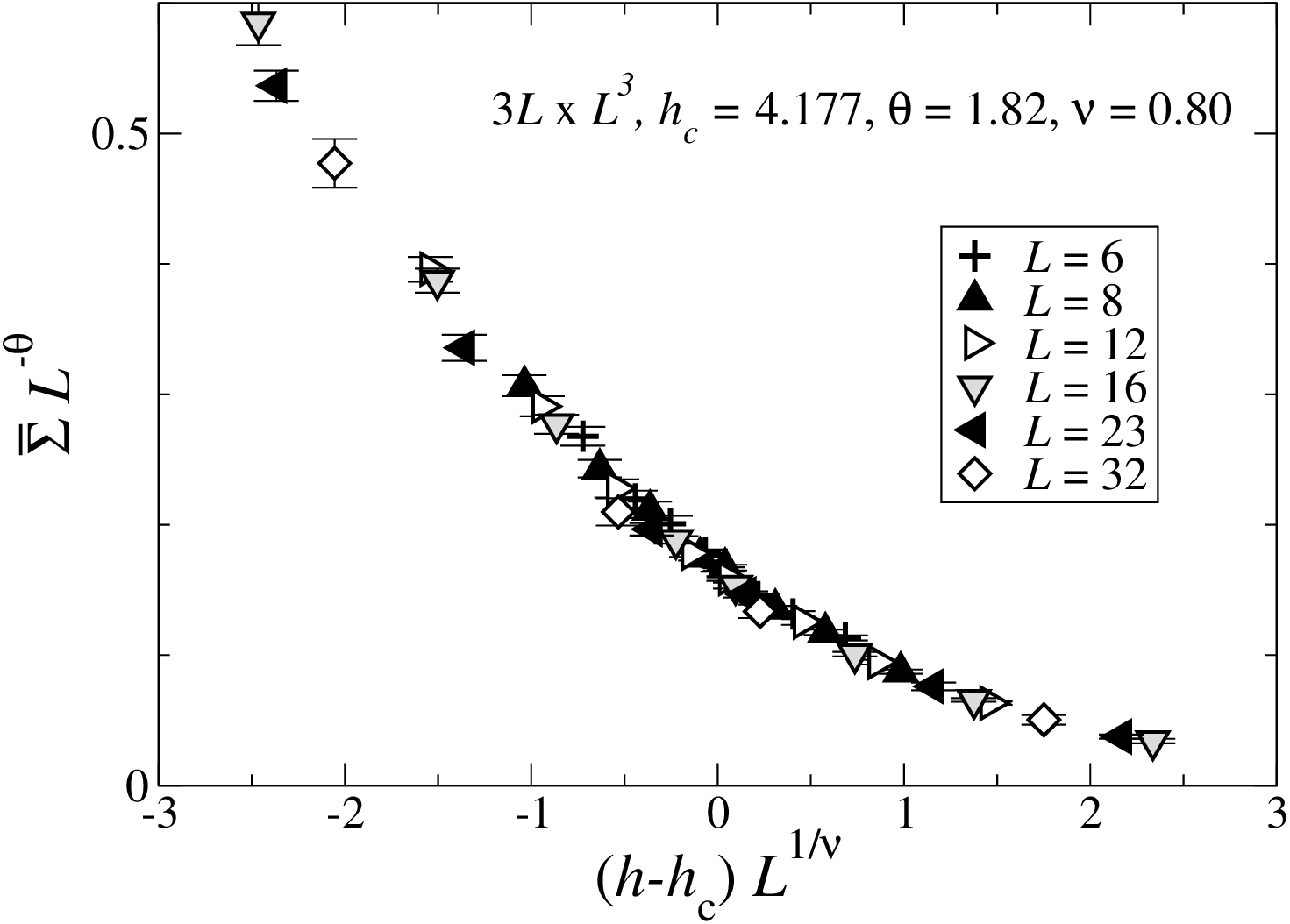}
\caption{
Scaling plot for the stiffness. The samples have a cross section
volume of $L^3$ with a distance of $3L$ between the controlled
faces. The scaled stiffness
$\Sigma L^{-\theta}$ is plotted vs.\ the scaled disorder
$(h-h_c)L^{1/\nu}$ for the values $\theta = 1.82$ and $\nu =0.80$.
}
\label{fig_stiffness}
\end{figure}

A scaling plot showing the collapse of the stiffness calculations for
samples of the same shape, $3L\times L^3$, is shown in \figref{fig_stiffness}.
Assuming the scaling form \eqref{eq_stiff}, a collapse to a single
function should be found when plotting $\overline{\Sigma}L^{-\theta}$
as a function of $(h-h_c)L^{1/\nu}$. This collapse is unreasonably
good (that is, is not too bad for $L=6$), using $h_c=4.177$,
$\theta=1.82$ and $\nu=0.80$. The computations strongly support the
picture of a second order transition with a value for $\theta$ obeying
the bounds \cite{FisherRFIM,VillainRFIM,SofferSchwartz}
\be
d/2-\beta/\nu\le\theta\le d/2.
\ee
When $d=3$, it has not been possible to determine whether $\theta\ne d/2$,
given that $\beta/\nu$ is so small. Here, given the larger value of
$\beta/\nu$, it is possible to discriminate between $\theta$
and $d/2$, with the result suggesting that $\theta < d/2$
(in addition, the result is consistent with
saturation of the lower bound.)

\subsection{Domain walls}

The calculations for the set of boundary conditions $++$, $--$, $+-$, and
$-+$ on the two opposite controlled faces (separated by $3L$) have also
been used to study the domain walls in the 4D RFIM.
Following the definitions of \refref{MiddletonFisherRFIM}, three definitions
of the domain wall are considered.
The first is found by comparing $++$ and $+-$ boundary conditions, with
the spins fixed to be $+$ on the left end in both cases. The set of spins
which is connected to the left end and fixed under both sets of boundary
conditions has an internal
boundary that intersects $W_s$ bonds. Assuming scaling
at $h_c$, the surface measure scales as
\be
\overline{W}_s \sim L^{d_s},
\ee
defining the domain-wall dimension $d_s$.
The next domain-wall measure exponent is found by comparing the $+-$
configuration with the $--$ and $++$ configurations. The number
of bonds which are
unsatisfied only with $+-$ boundary conditions gives a domain wall
measure $W_I$. Under $--$ and $++$ boundary conditions, there are
unsatisfied bonds due to frozen spin regions, where the random field
is strong enough to fix the spins under all boundary conditions. The
unsatisfied bonds with either of these two boundary conditions are not
counted as part of the $+-$ domain wall under this definition. The only
broken bonds which are counted as part of $W_I$ are those broken
due to the twisted boundary conditions.
This measure similarly defines an incommensurate surface exponent by
\be
\overline{W}_I \sim L^{d_I}.
\ee
The third definition of the effect of boundary conditions is given
by the bond or exchange part of the stiffness,
\be
\Sigma_J = (E_{+-}^J+E_{-+}^J-E_{++}^J-E_{--}^J)/2,
\ee
where $E_J=J\sum_{\left<ij\right>} s_is_j$.
This count includes some broken bonds with negative sign
and is influenced by frozen islands.
The ``dimension'' $d_J$ is then
\be
\overline{\Sigma}_J \sim L^{d_J}.
\ee
As $\overline{\Sigma}_J$ is the derivative of $\Sigma$ with respect
to $J$, thermodynamic relations\cite{MiddletonFisherRFIM} imply that
\be
d_J = \theta + \frac{1}{\nu}.
\label{dJscale}
\ee

\begin{figure}
\centering
\includegraphics[width=8.5cm]
{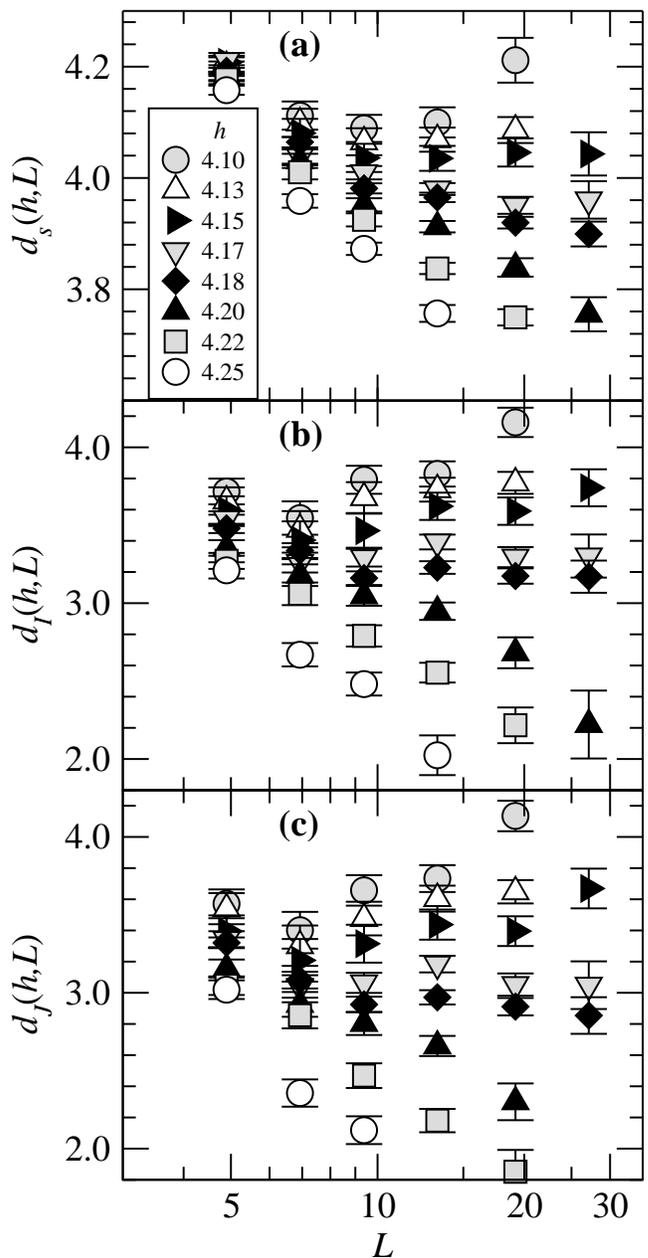}
\caption{Estimate of the dimensions (a) $d_s$, (b) $d_I$, and (c) $d_J$
obtained from the discrete logarithmic derivatives of
wall area $\overline{W}_s$,
the number of bonds $\overline{W}_I$ created by twisted BC's
relative to uniform sign BC's, and the exchange stiffness
$\overline{\Sigma}_J$ with respect to $L$ for several $h$. These plots
are used to infer $d_s = 3.94 \pm 0.06$,
$d_I = 3.20 \pm 0.12$ and $d_J = 2.94 \pm 0.12$.
}
\label{fig_dims}
\end{figure}

The values of $d_s$, $d_I$ and $d_J$ were estimated by 
taking the discrete logarithmic derivatives of $\overline{W}_{s}$,
$\overline{W}_I$ and $\overline{\Sigma}_J$,
\be
d_{s,I,J}(\sqrt{L_1L_2}) = \ln[W(L_2)/W(L_1)]/\ln(L_2/L_1),
\ee
with $W$ being one of the measures of the domain wall.
The results are plotted in \figref{fig_dims}.

One of the more striking differences between the 3D and 4D calculations
is that, while $d_s\ne d$ in 3D,
when $d=4$ the value of $d_s$ is consistent with the relation
\be
d_s=d.
\ee
Thus, the domain wall defined by the surface of connected fixed spins
anchored at the fixed end of the sample has dimension consistent
with the spatial dimension. This surface, the
internal boundary between flipped and fixed spins, appears to be
space-filling.
Additionally,
the estimated value of $d_I=3.20\pm 0.12$ is clearly
distinct from $d_s$, in contrast with the near equality seen in
3D.\cite{MiddletonFisherRFIM}
These differences
will be addressed in more detail in \secref{sec_perc},
when examining the frozen spins for $h<h_c$.

\subsection{Domain walls and scaling}

The value computed here for $d_J$, $d_J=2.94\pm 0.12$, is just
consistent with \eqref{dJscale}. As the derivation
of \eqref{dJscale} is quite robust, this consistency should
not be surprising. Though the arguments are apparently sound, the
conjecture made originally for 3D,\cite{MiddletonFisherRFIM} namely
that
\be
d_I=d_J+\beta/\nu,
\ee
has a non-rigorous derivation,
especially as in the above form, $d_s$ in the
original version has been replaced by $d_I$, which here more clearly
reflects the measure of the domain walls induced by boundary condition
changes. However, this scaling relation is easily consistent with the
computed values of the domain wall exponents and $\beta/\nu$.

The value found here for
$(\alpha-1)/\nu$ also satisfies the relationship
$d_I-d-\beta/\nu=(\alpha-1)/\nu$, which was used
in \refref{MiddletonFisherRFIM}, except for the replacement here of $d_s$
by $d_I$, motivated in the 4D RFIM by the more
natural definition of domain walls using $d_I$
and the spatial structure of frozen
spins.

\section{Frozen and minority spins}
\label{sec_perc}

The result that one measure of the domain wall dimension, $d_s$, is
near to the spatial dimension $d$ suggests that the picture of the
spin configurations must differ between the cases $d=3$ and
$d=4$. The picture of the configuration at the
transition in $d=3$ described
in \refref{MiddletonFisherRFIM} is that of nested domain walls, where
the domain walls are the boundaries separating connected sets of spins of
the same sign (see also \refref{VillainRFIM}.)
In this section, results are presented that
necessitate a different
picture in $d=4$, due to the percolation of
minority spins for $h$ less than the
critical disorder $h_c$.
(These results should be compared with those for the 3D
RFIM presented in
\refref{EsserNowakUsadel}, which support the existence of two
interpenetrating spanning domains in the 3D RFIM for $h>h_c$, and
those of \refref{SPAperc}, where the surprising
claim is made that there is a second
critical $h_p > h_c$ where there is first simultaneous spanning
by up and down spin clusters.)
The percolation of minority spins in $d=4$
for $h<h_c$ makes the identification of domain
walls with connected sets of uniform spins problematic.

Given a disorder realization $\{h_i\}$, there are two natural sets of
spins to consider when defining domain walls and percolation clusters.
The minority spins are simply those that have spin opposite to the mean
magnetization. The fraction of spins that fall into this category
is $(1-|m|)/2$. Frozen spins are those that are invariant under
all boundary conditions. These
spins are minority spins under either all up or all down boundary conditions,
so that the fraction of frozen spins is $1-|m|$, when $L\gg \xi$.
If either minority or frozen spins were distributed independently in
space, the clustering of these spins would map directly onto simple
percolation. As there are strong interactions between these spins and
the boundaries between them are related to domain walls, the percolation
is not simple, on short length scales. For $h<h_c$, the correlations
should vanish in the limit of separations much greater than $\xi$.

The clustering and percolation behavior of these spins can be directly
studied to learn more about the domain walls. From ground
state configurations for $L=8$ through $L=64$, computed both for all up
and all down boundary spins, the frozen and minority spin sets were
identified. These sets can be studied directly or in a coarse grained
sense. (Coarse grained spin blocks were determined by whether the
minority or frozen spins were a majority of the block, with
ties randomly broken.) The spanning clusters were defined as those that
connected two opposite faces of the hypercubic sample.

\begin{figure}
\centering
\includegraphics[width=8.5cm]
{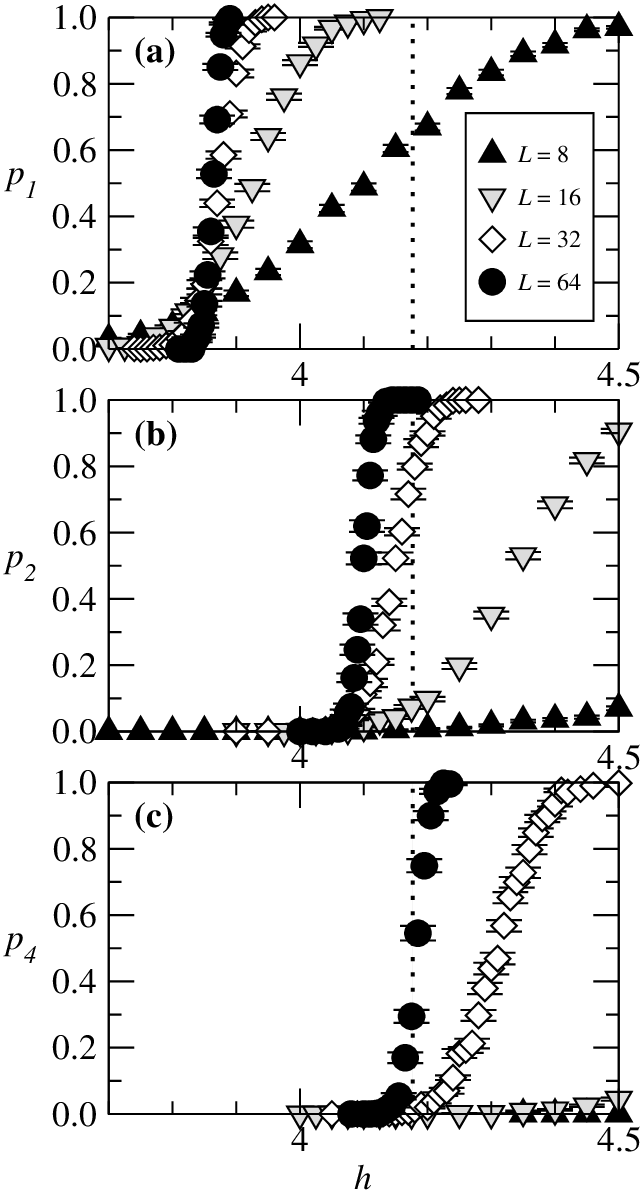}
\caption{Plot of the percolation probability for minority spins as
a function of disorder strength $h$, for $L=8,16,32,64$.
The dashed line indicates $h=h_c$.
(a) Percolation probability $p_1$ for minority spins. The percolation
probability approaches unity for large systems for $h_p^f = 3.875\pm0.005$.
(b) Percolation probability $p_2$ for minority block spins of size $2^4$.
The percolation threshold is closer to $h_c$.
(c) Percolation probability $p_4$ for minority block spins of size $4^4$.}
\label{fig_percb}
\end{figure}

Fig.\ \ref{fig_percb} is a plot of the percolation probability
$p_b$ (i.e., the probability of at least one spanning cluster)
of minority spins on scales $b=1,2,4$ as a function of $h$.
From this plot, an extrapolation of the curve crossings to large $L$
suggests that the minority spins percolate in the infinite-volume limit
at $h_p^m=3.850\pm 0.005$.
Plots for the frozen spins are qualitatively similar, with a lower percolation
threshold of $h_p^f=3.680\pm 0.005$.
In each case, the number of spanning clusters peaks near $h_p^{m,f}$,
with the peak number increasing with $L$.
The percolation point tends toward $h_c$
as the scale $b$ increases,
consistent with a scaling toward the ferromagnetic
state of uniform magnetization for $h<h_c$.

Implications for the simple domain wall picture in $d=4$ follow
directly. As the minority spins percolate in the
ferromagnetic phase, $h<h_c$, the boundaries
of connected sets of same-sign spins
are space filling.
The definition of domain walls using simply these
connected sets is thus not clearly informative about the effect of boundary
condition changes. The surfaces defined by the incongruent bonds are
more useful in understanding the domain walls. These are the surfaces that
separate the two ferromagnetic ground states; the frozen spins make a
space-filling background that is common to both states, even for $h<h_c$.
(In three dimensions, {\em at} $h=h_c$, there is a fractal 
set of spins that can be controlled by the boundary conditions.)
In addition, the relationship in 3D between $\beta/\nu$ and $\rho_\infty$,
the probability of crossing a domain wall per factor of $e$ in length
scale, would be much more difficult to investigate in 4D,
as the domain walls are not readily identifiable.

\section{States}
\label{sec_states}

In earlier sections, it has been (sometimes implicitly) assumed that
the transition in the 4D RFIM is consistent with the simple picture of
a ferromagnetic to paramagnetic transition, with the sign of the
magnetization in
the ferromagnetic state dependent on boundary conditions and the
spin configuration independent of boundary condition in the paramagnetic
state, far from the boundaries.
This assumption is examined in this section.
The approach is inspired in large part by
analytic work.\cite{FisherHuseStates,NewmanStein}
A discussion of the numerical
study of the nature of the thermodynamic
states is presented in \refref{AAMstates}
and the applications to the
3D RFIM can be found in \refref{MiddletonFisherRFIM}.
In summary, one test of the number of states in the thermodynamic
limit is to determine the correlation functions (in this case, the
ground state) in the interior of the 
sample under several different boundary conditions.
For a small number (one or two for Ising models) of thermodynamic
limits, there will be a small number of interior configurations.
The probability that the interior of the ground states will differ
from one of the large volume limit configurations decreases as
a power law dependent on the dimension of the domain wall.

The degeneracy of the ground state was directly addressed for
the 4D RFIM by studying
the effect of changing boundary conditions on the ground state spin
configuration in the interior of the sample. In particular, the
periodic ($P$), all spins up ($+$), all spins down ($-$), and open
($O$) boundary conditions were compared. As the 4D computations are much
more time consuming, the comparison between the ground states of
a system and a smaller subsystem, each with open boundary conditions, was
not extensively studied, as it was in the 3D RFIM.

\begin{figure}
\centering
\includegraphics[width=8.5cm]
{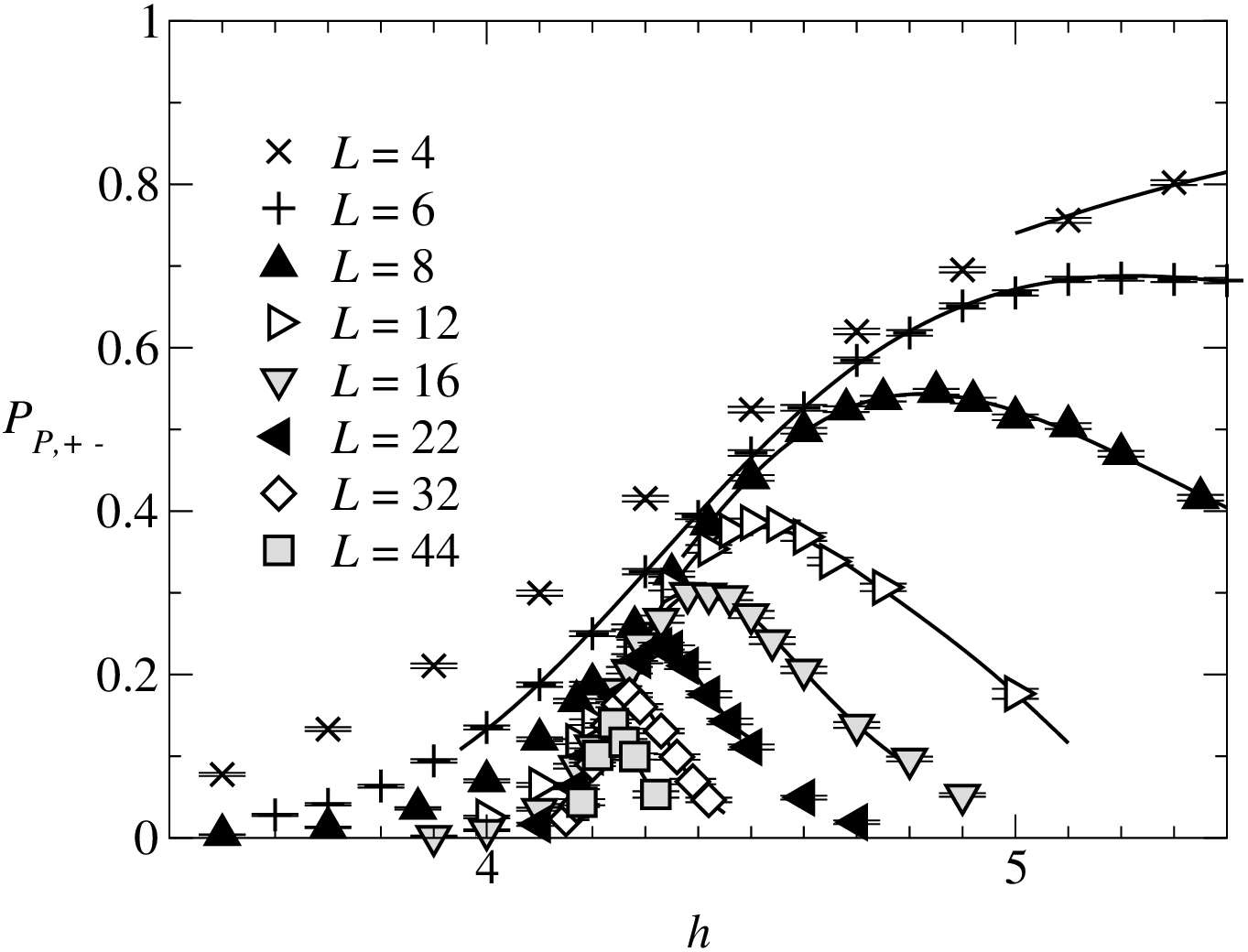}
\caption{Plot of $P_{P,+-}(2,h,L)$, the probability that, at
given sample size $L$ and disorder $h$, the ground state with periodic
boundary conditions will differ from ground state configurations for
both fixed $+$ and $-$ boundary conditions,
in a volume of size $2^4$ in the center
of the volume $L^4$. Note that extrapolating to $L\rightarrow\infty$ suggests
that $P_{P,+-}(2,h,L)\rightarrow 0$ for all $h$. The solid lines
are least-squares fits to quartics in $\ln(P)$.
}
\label{fig_P}
\end{figure}

\begin{figure}
\centering
\includegraphics[width=8.5cm]
{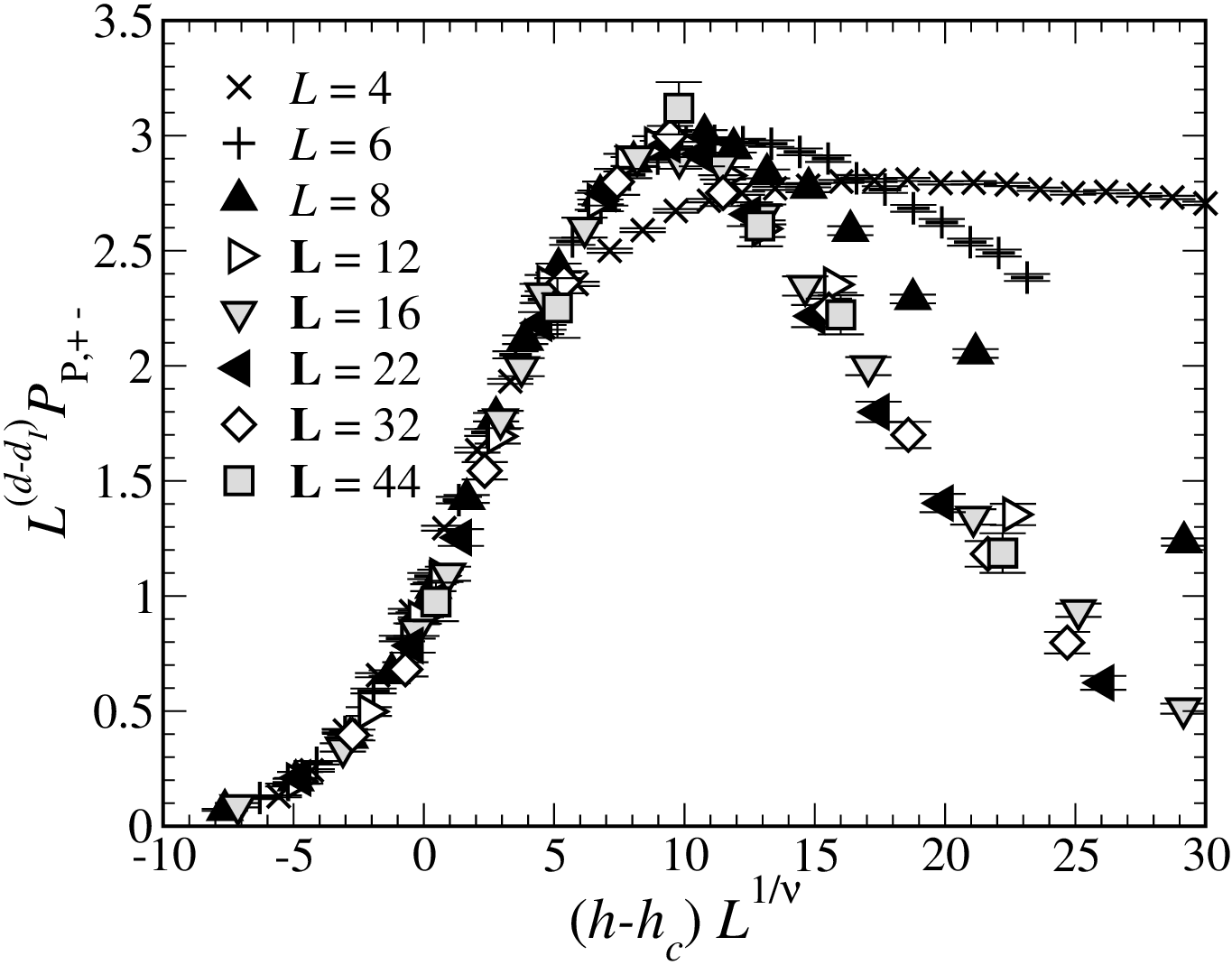}
\caption{
Scaling plot for $P_{P,+-}(2,h,L)$, using the scaled probability
$P_{P,+-}L^{d-d_I}$ and scaled disorder $(h - h_c)L^{1/\nu}$, with
values $d_I=3.2$ and $\nu = 0.8$. The scaling collapse suggests that the
changes in boundary conditions typically introduce a finite
number of domain walls (say, one of size $L$) with
$\sim L^{d_I}$ bonds at $h = h_c$.
}
\label{fig_Pscale}
\end{figure}

A summary of the results for comparisons between $P$ and $+$/$-$
is presented in Figs.\ \ref{fig_P} and \ref{fig_Pscale}.
The results for $O$ vs.\ $+$/$-$ are quite similar.
The plots show the (scaled) probability $P_{P,+-}(2,h,L)$, for
a given $h$ and $L$, that the ground-state $P$ configuration is distinct from
both the ground-state
$+$ and $-$ configurations in the central volume of size
$2^4$.
As $L$ increases, this probability decreases toward zero at all $h$.
This suggests that in large samples, the interior configuration for
a number of boundary conditions (including periodic and open) can
be found by imposing either $+$ or $-$ boundary conditions.
It was also found, as in the 3D case, that for $h>h_c$, as $L$ increases,
the interior configurations for the $+$ and $-$ boundary conditions
become identical with unit probability.
Together, assuming the extrapolation to large $L$ is correct,
these results show the existence of a single
state for $h>h_c$ (for if the interior configuration differs
between
{\em any} two boundary conditions, it must differ
between $+$ and $-$) and
strongly suggest the existence of only two states for $h<h_c$.

The scaling of $P_{P,+-}$ is consistent with previous work on
disordered models.\cite{AAMstates,PalassiniYoung2d} As a function
of the scaled disorder $(h-h_c)L^{1/\nu}$, the function approaches
a single curve when the probability $P_{P,+-}$ is scaled by $L^{d-d_I}$.
This scaling results from assuming that
the number of large (size $L$) domain walls
induced by generic boundary condition
changes is constant as $L\rightarrow\infty$.
This assumption is consistent with the observation that $d_I > d-1$.
The chance that an
interior volume of fixed linear size $w$
intersects a domain wall is expected to behave as
$(w/L)^{d-d_I}$. The clean collapse of the data for larger system
sizes, using values determined from magnetization and domain wall
measurements, lend quantitative support to this picture.

\begin{figure}
\centering
\includegraphics[width=8.5cm]
{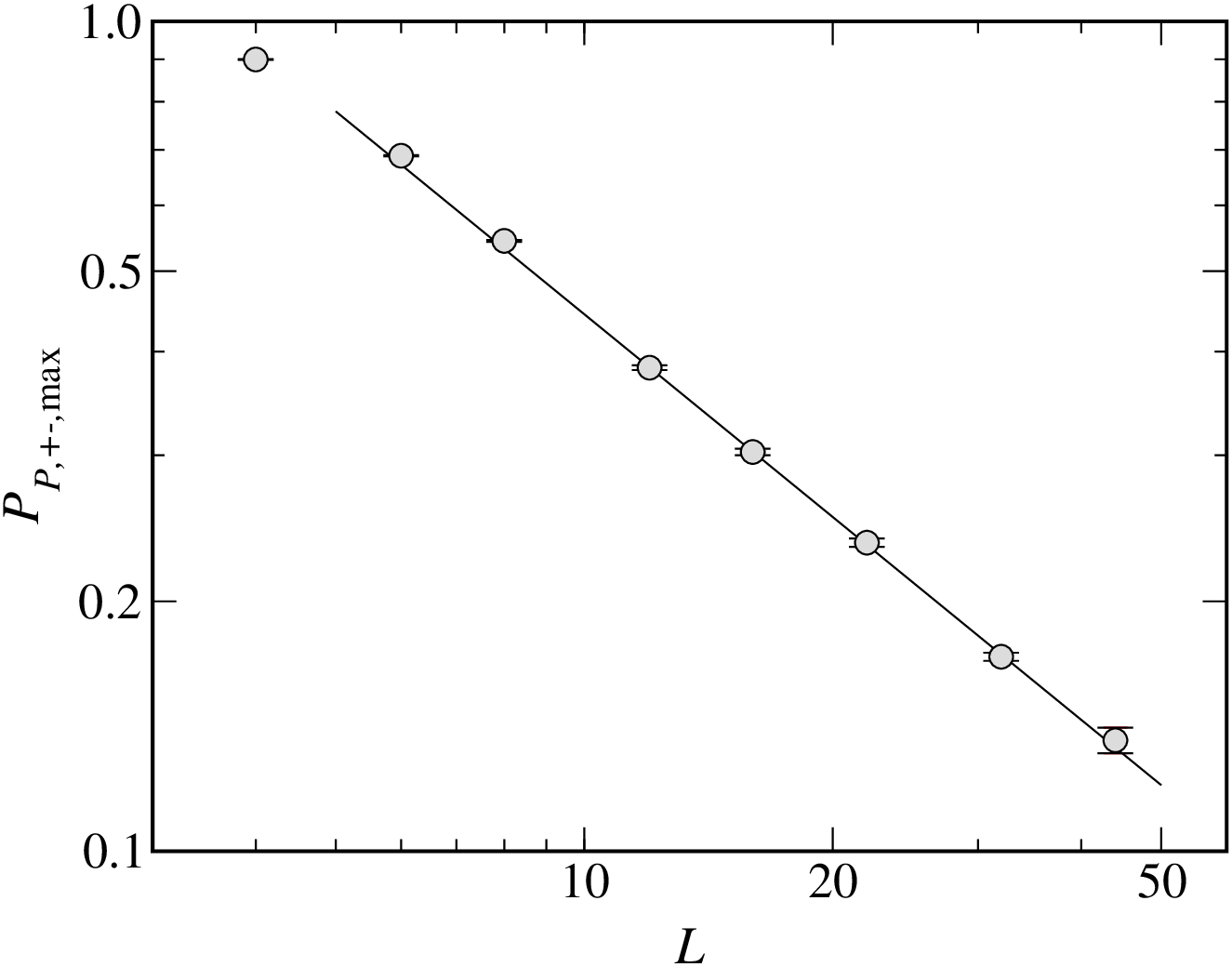}
\caption{
Plot of the dependence of the maximum of $P_{P,+-}(2,h,L)$ over $h$
on the system size $L$. The solid line is a fit
to $P_{P,+-}^{\rm max} \sim L^{-(d-d_I)}$,
with $d-d_I=0.81\pm0.02$.
}
\label{fig_Pmax}
\end{figure}

The dimension $d_I$ could be alternatively deduced from this
data. \figref{fig_Pmax} shows the dependence of the peak value
of $P_{P,+-}$ on $L$. {\em Assuming} $d-d_I$ gives this slope, the
data for $L=12\rightarrow 44$ gives $d_I=3.19(2)$. This value could be
taken as the best one for $d_I$, but this is not done here and is instead
used as a confirmation of the scaling picture.

\begin{figure}
\centering
\includegraphics[width=8.0cm]
{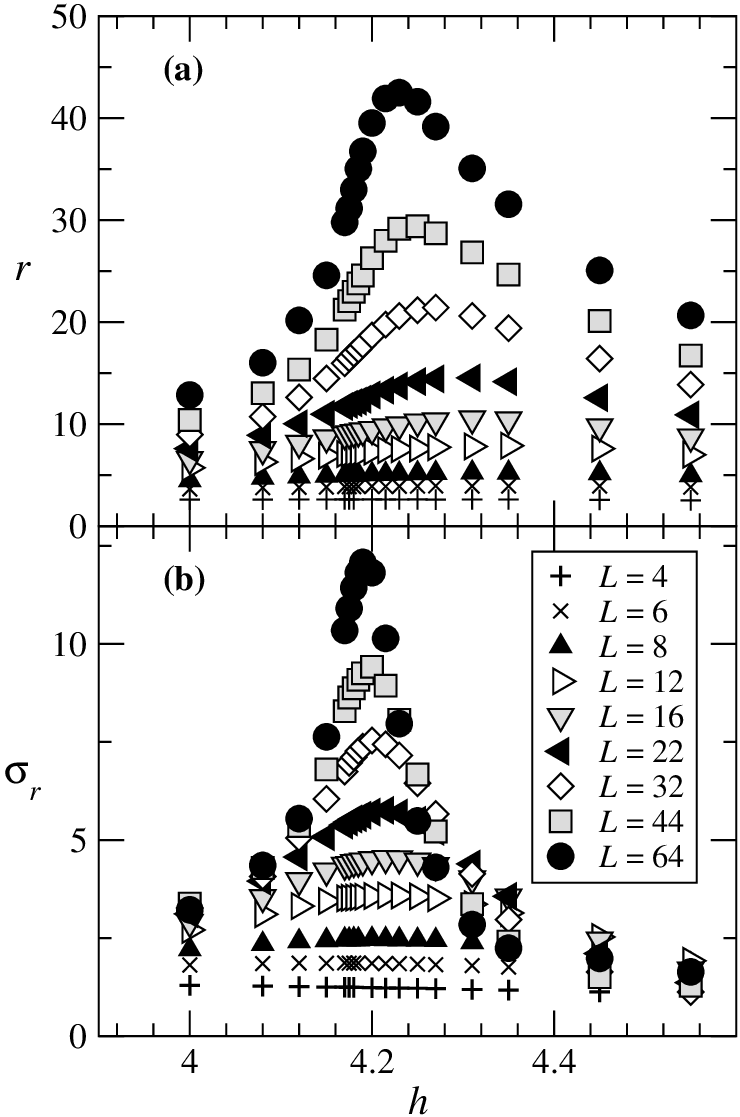}
\caption{(a) Plot of the number of relabel operations per spin,
$r$, carried out by the ground state algorithm
as a function
of $h$, for $L=4\ldots64$.
(b) Plot of the sample-to-sample fluctuations $\sigma_r$
in the number of relabel operations per spin. This quantity provides an
especially sharp and quickly diverging curve for estimating $h_c$.
}
\label{fig_relabel}
\end{figure}

\begin{figure}
\centering
\includegraphics[width=8.0cm]
{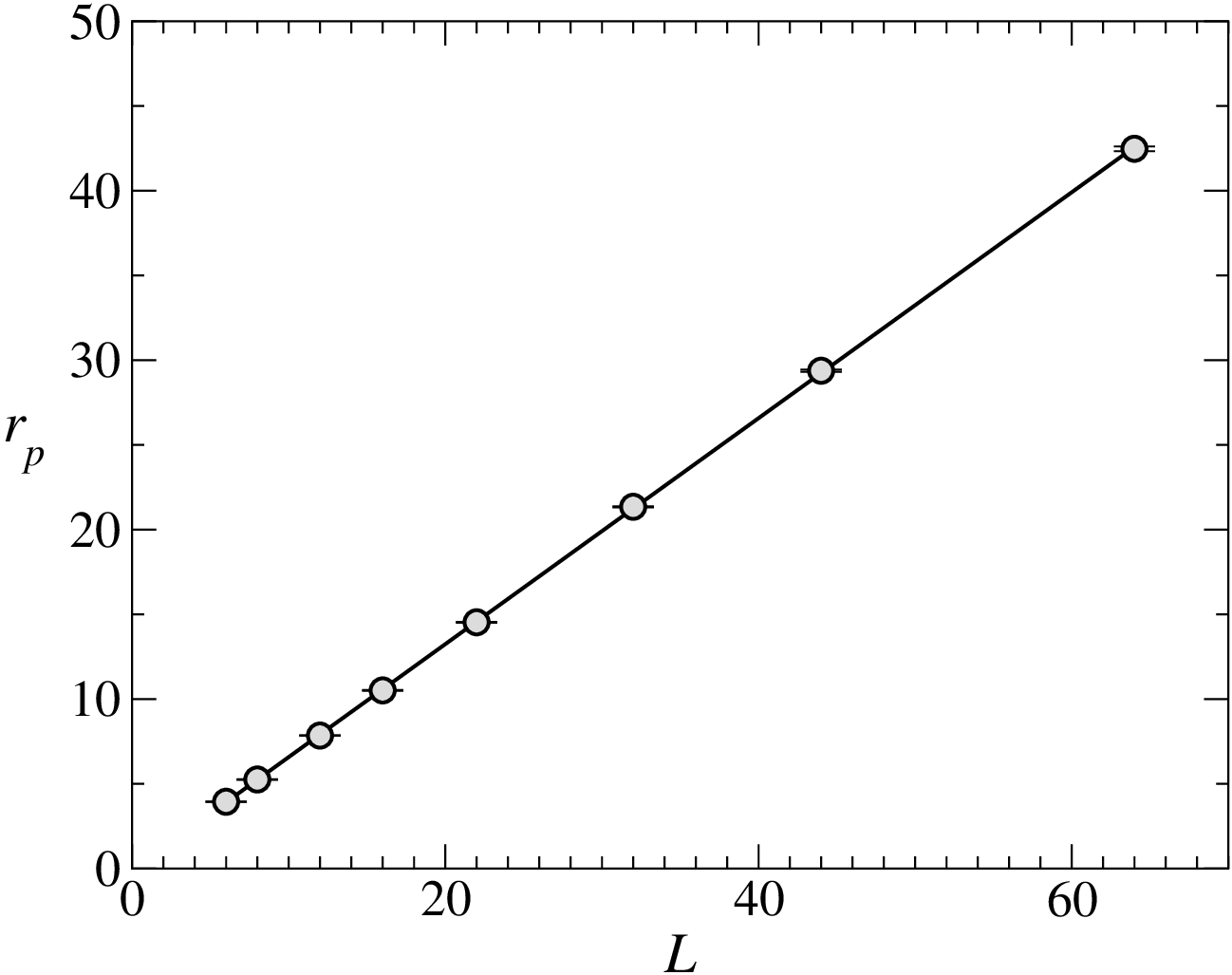}
\caption{
Linear plot of the height of the peak $r_p(L)$ in $r$ vs.\ $h$.
The value of $r_p$ scales almost exactly linearly with $L$.
}
\label{fig_relabelmax}
\end{figure}

\section{Algorithmic slowing down}

In the 3D RFIM, it was found that the number of operations carried out
by the ground state algorithm diverged near the ferromagnetic-paramagnetic
transition. In this section, similar results are presented for the 4D RFIM.
The scaling results here have greater
accuracy near the transition. The scaling is found to
be quite consistent with
the heuristic picture presented in \refref{AAMslowing}.

The key quantitative
relation to be elucidated is that between the number of primitive
operations carried out to find the ground state and the physical
understanding of the phase transition and correlation volumes.
In \refref{AAMslowing}, it was argued that the time per spin to find the
ground state in the RFIM is directly proportional to the linear size
$L$ near the transition. This results from the nature of the push-relabel
algorithm\cite{ADMR,MiddletonFisherRFIM} used, which
efficiently constructs a ``height'' field over the lattice that
guides the ``flow'' (corresponding to the redistribution of
excess ``fluid'' or external magnetic field $h_i$.)

The relabel operation is one of the primitive operations carried out
during the convergence of the algorithm to the physical ground state.
In \figref{fig_relabel}, the sample average $r$ of
the number of relabel operations per spin
and the sample-to-sample
fluctuations in $r$, $\sigma_r$,
are plotted as a function of $h$ for different $L$.
(Results for the number of the other primitive
operations, the push operations, are quite similar.)
There is clearly a peak in both quantities near $h_c$, with the peak
in the sample-to-sample fluctuations being more sharply peaked, relative
to the non-critical contribution \cite{noncrit}.

The $r(h)$ curves were fit with fourth-order polynomials to
extract the peak value $r_p(L)$. The plot of this quantity is shown
in \figref{fig_relabelmax}. A linear fit is shown, which is remarkably
consistent with the data over a wide range of $L$. The result is
in agreement with the arguments of \refref{AAMslowing} and supports the
relationship between the physical correlation length and the evolution
of the algorithm.

\section{Summary}

By computing the ground state for a large number of samples of
volume up to $L^4=64^4$, the quantitative and qualitative thermodynamic
properties of the 4D RFIM have been studied. The derived exponents
satisfy the conventional scaling relations.
The values of the exponents and location of the transition are consistent
with, but are based on larger systems than,
the results for the Gaussian
4D RFIM published in Refs.\ \onlinecite{Swiftetal,Hartmann4d}.
Note that, as in previous work, the value for $\nu$ is the least
certain and that errors in $\nu$ propagate to estimates of $\beta$ and
$\alpha$.
The picture of a single transition from a nearly-two-fold-degenerate
ferromagnetic state to a single paramagnetic state is confirmed by
comparing ground states with varying boundary conditions.
Strong evidence is presented that the picture of domain walls
developed\cite{MiddletonFisherRFIM} for the 3D RFIM must be modified
to describe the 4D RFIM. In particular, the percolation of frozen and
minority spins within the ferromagnetic phase implies that the
sets of connected same-sign spins are not the boundaries of domain
walls. The empirical running times for the ground state algorithm
peak near the phase transition in a manner consistent with
previous descriptions,\cite{MiddletonFisherRFIM,AAMslowing}
with the peak running time per spin apparently proportional to
the linear system size. The algorithmic running times provide a check
on the location of the transition and the scaling exponent $\nu$.

I would like to thank Daniel Fisher for discussions and
Alexandar Hartmann for communicating related work prior to its submission.
This material is based on work supported by the
National Science Foundation under Grant No. DMR-0109164.


\begin{thebibliography}{xx}
\bibitem{noncrit}
The non-critical background to $r$ or $\sigma_r$
may, however, have a magnitude that
slowly diverges with $L$, as discussed in \refref{AAMslowing}.
\end{thebibliography}
\end{document}